\documentclass[12pt]{article}
\usepackage{amsmath,amssymb}
\usepackage{graphicx}
\usepackage{overpic}
\usepackage{dsfont}
\usepackage{times}
\usepackage{bm}
\usepackage[plain,noend]{algorithm2e}
\usepackage[hang]{subfigure}
\usepackage{natbib}

\newcommand{\blind}{0}

\newtheorem{prop}{Proposition}
\newtheorem{thm}{Theorem}
\newtheorem{coro}{Corollary}

\newcommand\mA{\mathcal{A}}
\newcommand\ols{\textnormal{ols}}

\def\T{{\mathrm{\scriptscriptstyle T} }}
\newcounter{rmk}
\newcommand\rmk[1]{\vspace*{1mm} \par \stepcounter{rmk}{
R{\footnotesize EMARK} \thermk}. {#1}\vspace*{1mm}}

\begin{document}

\def\spacingset#1{\renewcommand{\baselinestretch}%
{#1}\small\normalsize} \spacingset{1}


\if0\blind
{
  \title{\bf MSP: A Multi-step Screening Procedure\\ for Sparse Recovery}
  \author{Yuehan Yang\thanks{
    The authors gratefully acknowledge \textit{the National Natural Science Foundation of China (Grant No. 11671059) and the Program for Innovation Research in Central University of Finance and Economics.}}\hspace{.2cm}\\
    Department of Statistics, Central University of Finance and Economics,\\
    Ji Zhu \\
    Department of Statistics, University of Michigan\\
    and \\
    Edward I. George\\
    Department of Statistics, The Wharton School, University of Pennsylvania
    }
  \maketitle
} \fi

\if1\blind
{
  \bigskip
  \bigskip
  \bigskip
  \begin{center}
    {\LARGE\bf MSP: A Multi-step Screening Procedure for Sparse Recovery}
\end{center}
  \medskip
} \fi

\bigskip
\begin{abstract}
We propose a Multi-step Screening Procedure (MSP) for the recovery of sparse linear models in high-dimensional data. This method is based on a repeated small penalty strategy that quickly converges to an estimate within a few iterations. Specifically, in each iteration, an adaptive lasso regression with a small penalty is fit within the reduced feature space obtained from the previous step, rendering its computational complexity roughly comparable with the Lasso. MSP is shown to select the true model under complex correlation structures among the predictors and response, even when the irrepresentable condition fails. Further, under suitable regularity conditions, MSP achieves the optimal minimax rate $(q \log n /n)^{1/2}$ for the upper bound of $l_2$-norm error.  Numerical comparisons show that the method works effectively both in model selection and estimation, and the MSP fitted model is stable over a range of small tuning parameter values, eliminating the need to choose the tuning parameter by cross-validation. We also apply MSP to financial data and show that MSP is successful in asset allocation selection.
\end{abstract}

\noindent%
\vfill

\newpage
\spacingset{1.5} 
\section{Introduction}
Sparse recovery is of paramount interest in high-dimensional statistical problems where many predictors are available yet the regression function is well approximated by a few relevant covariates. A seminal contribution to this endeavor, the Lasso \citep{tibshirani1996lasso} simultaneously performs model selection and parameter estimation through regularization with a convex penalty. Now widely used for sparse recovery in practice, further extensions of the Lasso have enhanced its applicability and offered some theoretical guarantees, for example, see \citet{efron2004lars,friedman2010regularization,meinshausen2006high,zhao2006lasso,zou2006adaptive,fan2010selective,hastie2015statistical,buhlmann2011statistics}.

Although convex regularization methods such as the Lasso are computationally attractive and enjoy great performance in prediction, they also lead to biased estimates and require rather restrictive conditions on the design matrix to obtain model selection consistency. Nonconvex penalization procedures such as SCAD \citep{fan2001variable}, MCP \citep{zhang2010mcp} and the Spike-and-Slab Lasso (SSL) \citep{rovckova2018spike} have been proposed to lessen the bias. Multi-step methods do this too, including \citet{zhangtong2010} who proposed the Capped$-l_1$ regularization, leading to a multi-step convex relaxation scheme which is shown to obtain the correct feature set after a certain number of iterations. \citet{zou2008one} proposed a unified algorithm based on the local linear approximation (LLA) for maximizing the penalized likelihood, presenting a one-step low-dimensional asymptotic analysis for justification. \citet{fan2014strong} provided a unified theory to show how to obtain the oracle solution via LLA. The theoretical properties of LLA highly rely on the initial estimates. \citet{buhlmann2008discussion} proposed a method called multi-step adaptive lasso (MSA-Lasso), which updates the adaptive weights and re-estimates {\it the entire set of regression coefficients} at each iteration until convergence. \citet{huang2012concave} showed that, under certain conditions, the multi-step framework can improve the solution quality. Further work focusing on multi-step methods includes \citet{liu2016concave,wang2013concave,zhang2012concave}.

In spite of the fact that most nonconvex penalties do not require the irrepresentable condition to achieve model selection consistency \citep{fan2001variable,zhang2010mcp}, identifying the relevant predictors in the presence of highly collinear predictors may still present numerical challenges, as shown in Sections 4 and 5.  Indeed, nonconvex penalties can introduce numerical difficulties in fitting models, becoming less computationally efficient than convex optimization problems.

The main thrust of this paper, is to propose a Multi-step Screening Procedure (MSP), a simple multi-step method with the following characteristics:
\begin{itemize}
\item[(1)] MSP applies a single small penalty parameter, which remains fixed throughout, to minimize bias at each iteration. At each step, the active set is shrunk by deleting the ``useless'' variables whose coefficients have been thresholded to 0. When dealing with high dimensional data, this strategy will start off with a large model with many possibly incorrect variables and iteratively distinguish the nonzeros from zeros.
\item[(2)] This backward deletion strategy of MSP significantly reduces the execution time of the multi-step method. As will be seen in simulations, the computational complexity of MSP is roughly comparable to the solution path of Lasso.
\item[(3)] With an inherently small estimation error bound, MSP successfully recovers the true underlying sparse model even when the irrepresentable condition is relaxed. Indeed, MSP remains effective even when the irrelevant variables are strongly correlated with the relevant variables. Note that although many nonconvex methods do not require restrictive conditions on the design matrix in theory, they may still have difficulty in selecting the right model with finite samples. MSP is much better able to deal with such data.
\item[(4)] It is seen in simulations that the MSP fitted model is stable over a range of small tuning parameter values, eliminating the need to choose the tuning parameter by cross-validation. The solution of this method is both sparse and stable.
\end{itemize}
This paper is organized as follows. Section 2 presents the method and discusses its relationship to other methods. Section 3 shows its theoretical properties. The simulations in Section 4 and application in Section 5 assess the performance of the proposed method and compare it with several existing methods. Technical details are provided in the Supplementary Material.

\section{Method}\label{method}
In this section, we present the details of the MSP algorithm and compare it with existing methods. We consider the linear regression problem:
\[y = X\beta +\epsilon,\]
where $y$ is an $n$ response vector, $X$ is an $n \times p$ matrix, $\beta$ is a vector of regression coefficients and $\epsilon$ is the error vector. We are particularly interested in the case where the number of parameters greatly exceeds the number of observations ($ n \ll p $). We consider the $q$-sparse model, where $\beta$ has at most $q$ nonzero elements. Components of the error vector $\epsilon$ are independently distributed from $N(0, \sigma^2)$. The data and coefficients are allowed to change as $n$ grows; meanwhile, $p$ and $q$ are allowed to grow with $n$. For notational simplicity, we do not index them with $n$.

Recall that the Lasso estimator \citep{tibshirani1996lasso} minimizes squared error loss regularized with the $l_1$-penalty. Compared to least squares, Lasso shrinks a particular set of coefficients to zero while shrinking the others towards zero. These two effects, model selection and shrinkage estimation, are controlled only by a single tuning parameter, leading to its well-known estimation bias. Although \citet{zhao2006lasso} and \citet{meinshausen2006high} proved that the Lasso is model selection consistent under an irrepresentable condition, the condition is, however, quite restrictive. To mitigate these drawbacks, we propose MSP with two goals in mind: 1) recovery of the true sparse model when the irrepresentable condition fails; and 2) ``almost unbiased estimation'' by lowering the influence of the shrinkage penalty.

The essential idea behind MSP is to provide more precise estimation through iterated penalization with a smaller tuning parameter that is less influential at each step.  More precisely, the MSP Algorithm proceeds as follows.

\begin{itemize}
\item Initialize $k=1$. Obtain a lasso solution $\hat \beta^{[1]}(\lambda_0)$:
\[\hat \beta^{[1]} := \arg \min \Big\{\dfrac{1}{2}\|y -  X \beta\|^2_2+ \lambda_0 \|\beta\|\Big\}\]
and let $\mA^{[1]}$ be the nonzero index set of $\hat \beta^{[1]}$, i.e. $\mA^{[1]} = \{j\in\{1,...,p\}: \hat \beta_j^{[1]} \neq 0\}$.
\item Repeat the following steps until convergence:
\begin{eqnarray*}
k &\longleftarrow& k+1, \\
\hat{\beta}^{[k]} &:=& \mathop{\arg\min}\limits_{\beta_{(\mA^{[k-1]})^c}=0}
\Big\{ \dfrac{1}{2}\|y -  X \beta\|^2_2+ \lambda \sum\limits_{ j \in \mA^{[k-1]}}|\beta_j/\hat \beta_j^{[k-1]}|\Big\},
\end{eqnarray*}
where the active set $\mA^{[k]}$ is updated in every step, i.e. $\mA^{[k]} = \{j\in \{1,...,p\}: \hat \beta^{[k]}_{j} \neq 0 \}$.
\end{itemize}
At convergence, denote the active set by $\mA$ and the solution by $\hat \beta$. Note that the active sets obtained during the iterations are nested, i.e.
\[\mA^{[1]} \supseteq \mA^{[2]} \supseteq \cdots \supseteq \mA^{[k]} \supseteq \cdots \supseteq \mA, \]
as in each iteration an adaptive lasso is fit using only the features selected by the previous step. This is key to control the computational time of the algorithm as well as to maintain a rather small tuning parameter $\lambda$. We will provide more details on the choice of this small tuning parameter in the theoretical results.

\begin{figure}[ht]
\centering
\subfigure[MSP]{
\includegraphics[width=.23\textwidth,height=.23\columnwidth]
{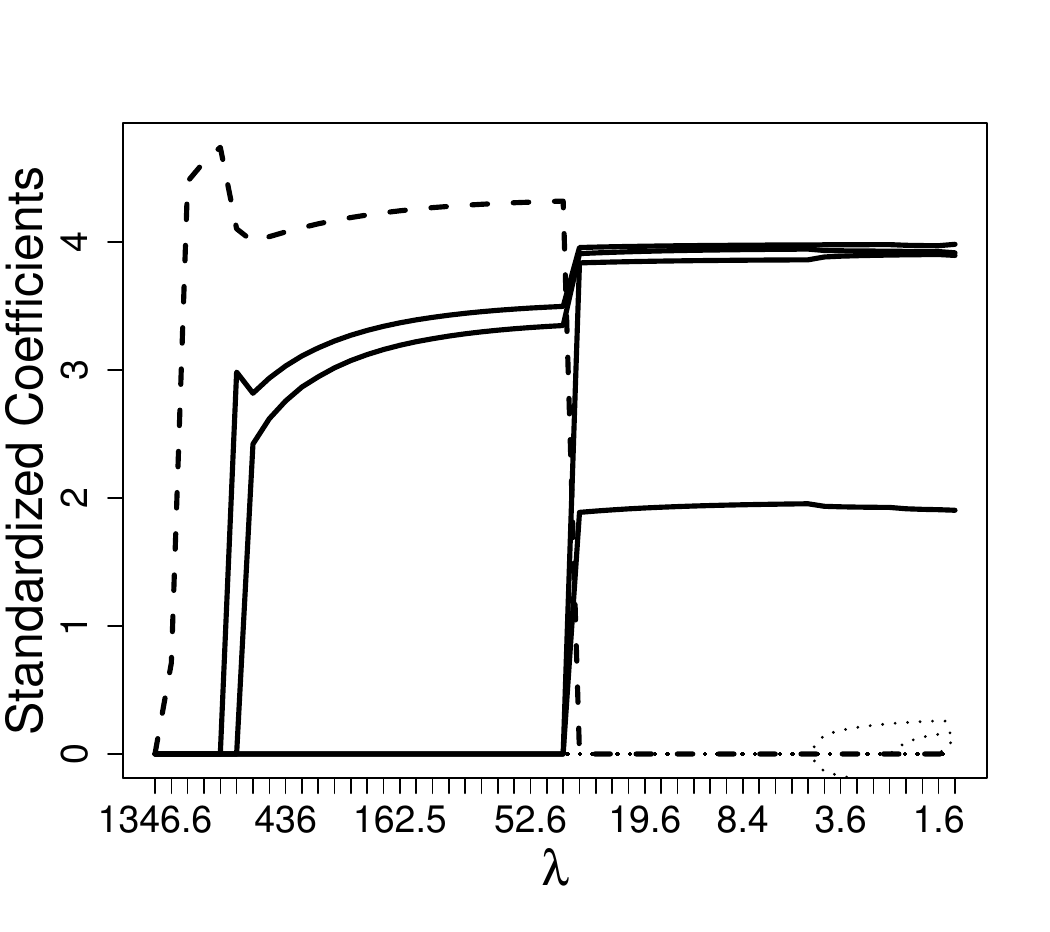}}
\subfigure[Capped$-l_1$]{
\includegraphics[width=.23\textwidth,height=.23\columnwidth]
{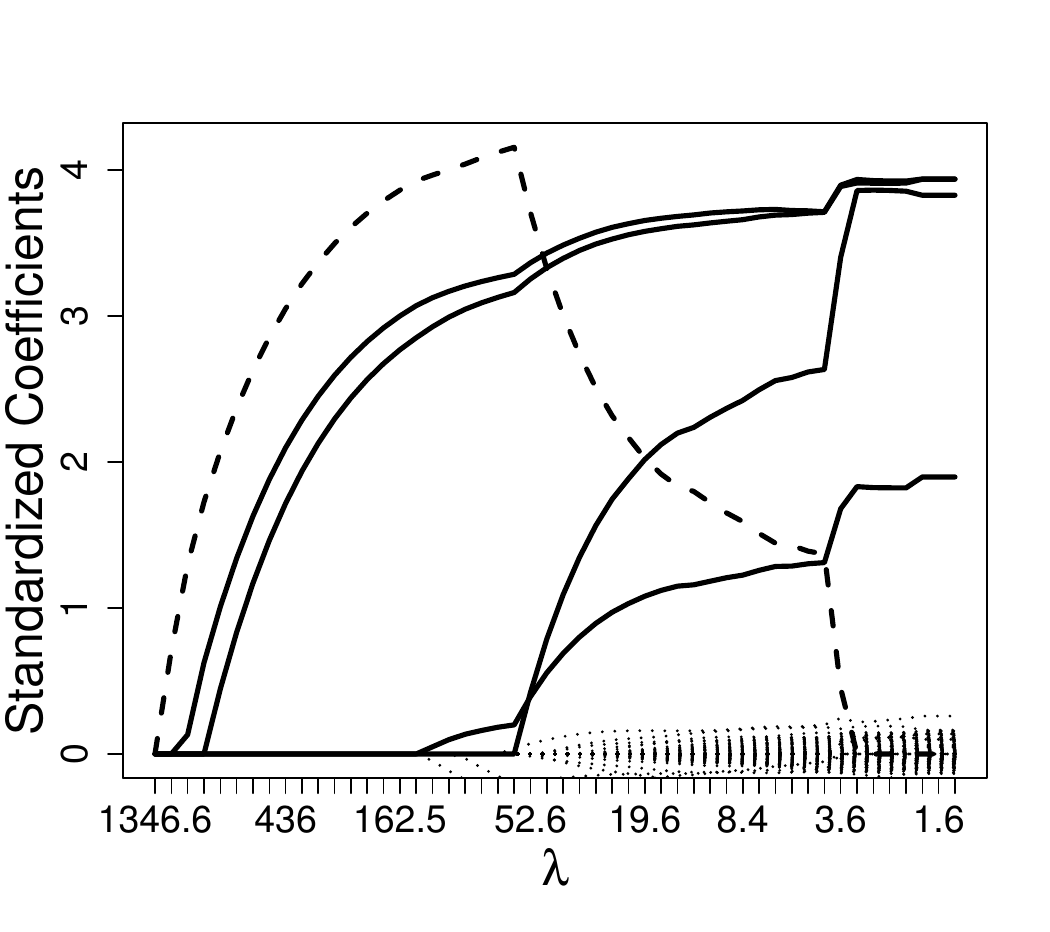}}
\subfigure[LLA]{
\includegraphics[width=.23\textwidth,height=.23\columnwidth]
{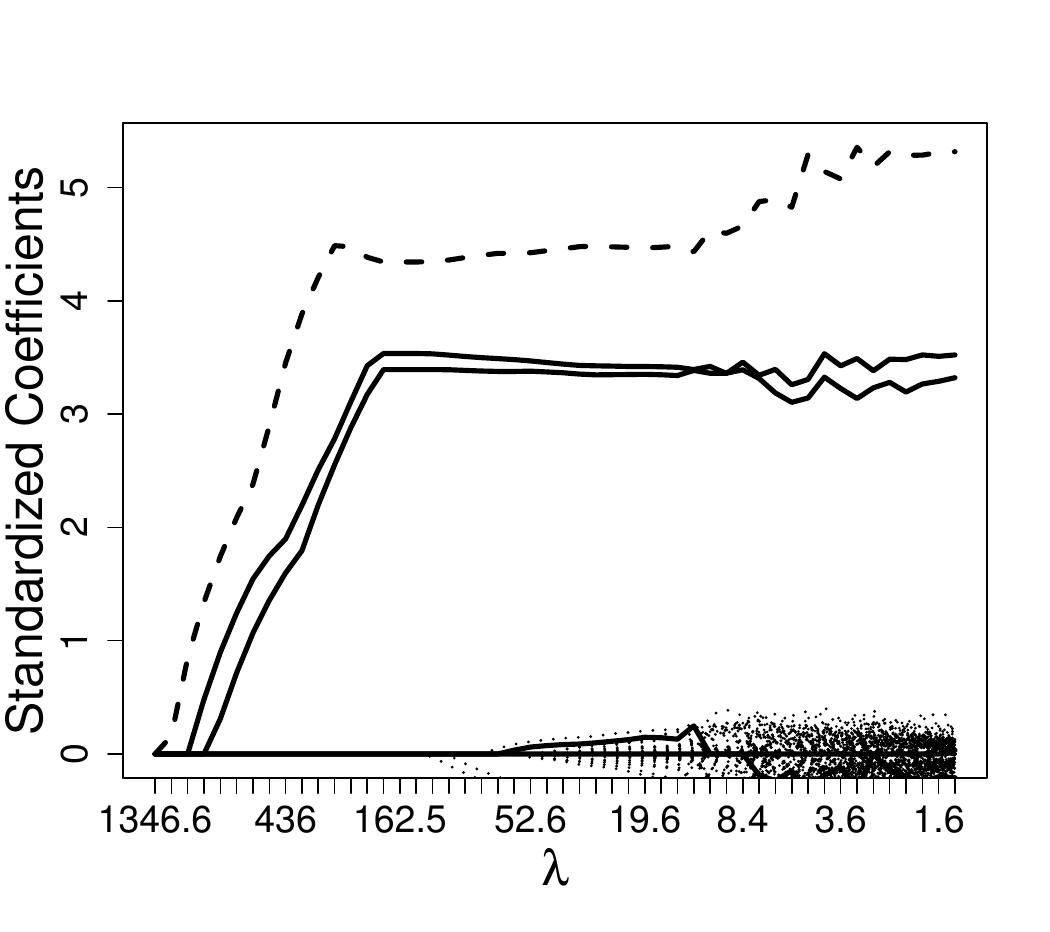}}
\subfigure[Adaptive Lasso]{
\includegraphics[width=.23\textwidth,height=.23\columnwidth]
{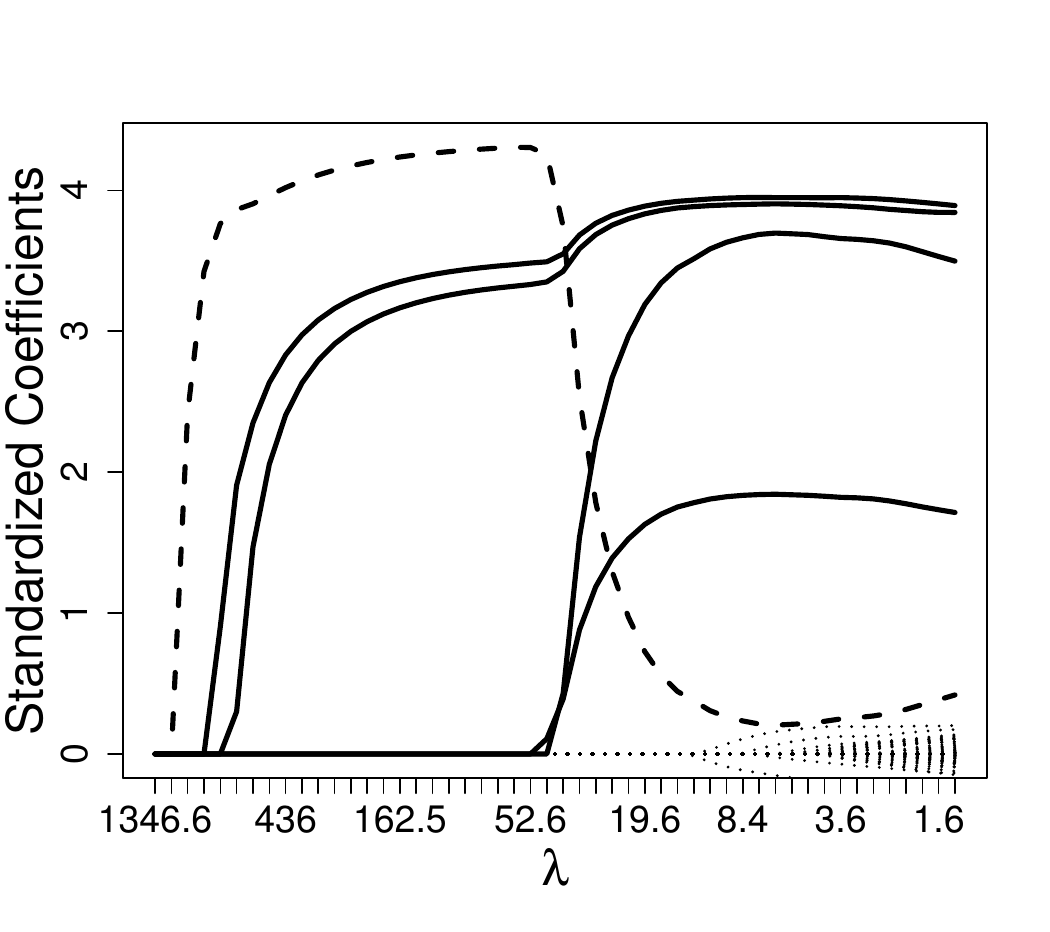}}
\subfigure[Lasso]{
\includegraphics[width=.23\textwidth,height=.23\columnwidth]
{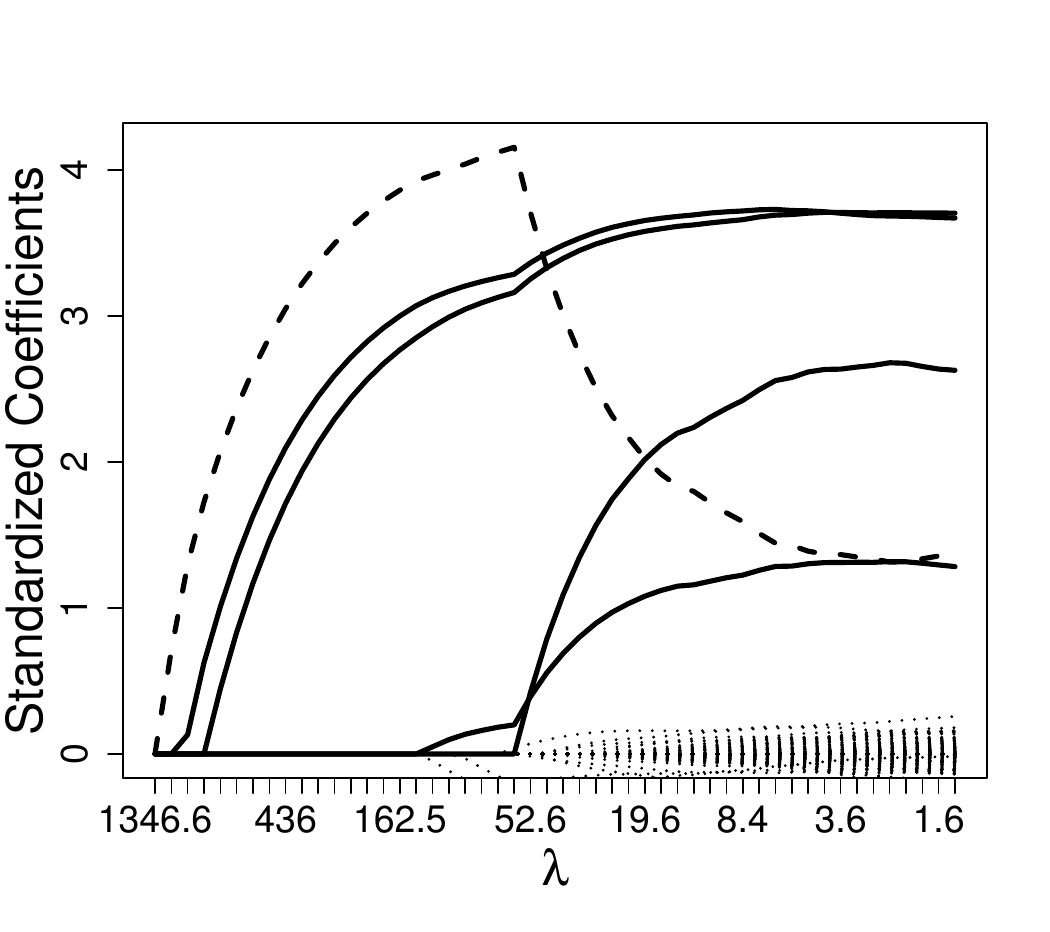}}
\subfigure[SCAD]{
\includegraphics[width=.23\textwidth,height=.23\columnwidth]
{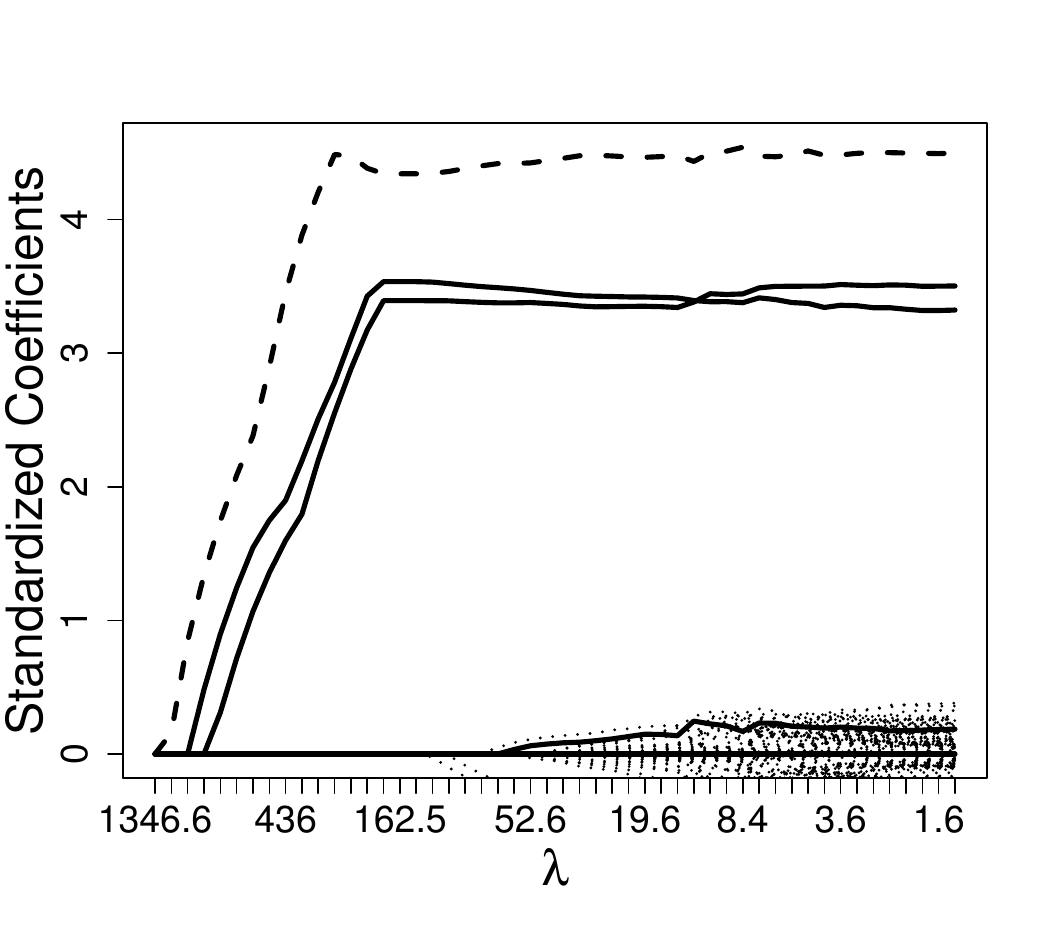}}
\subfigure[MCP]{
\includegraphics[width=.23\textwidth,height=.23\columnwidth]
{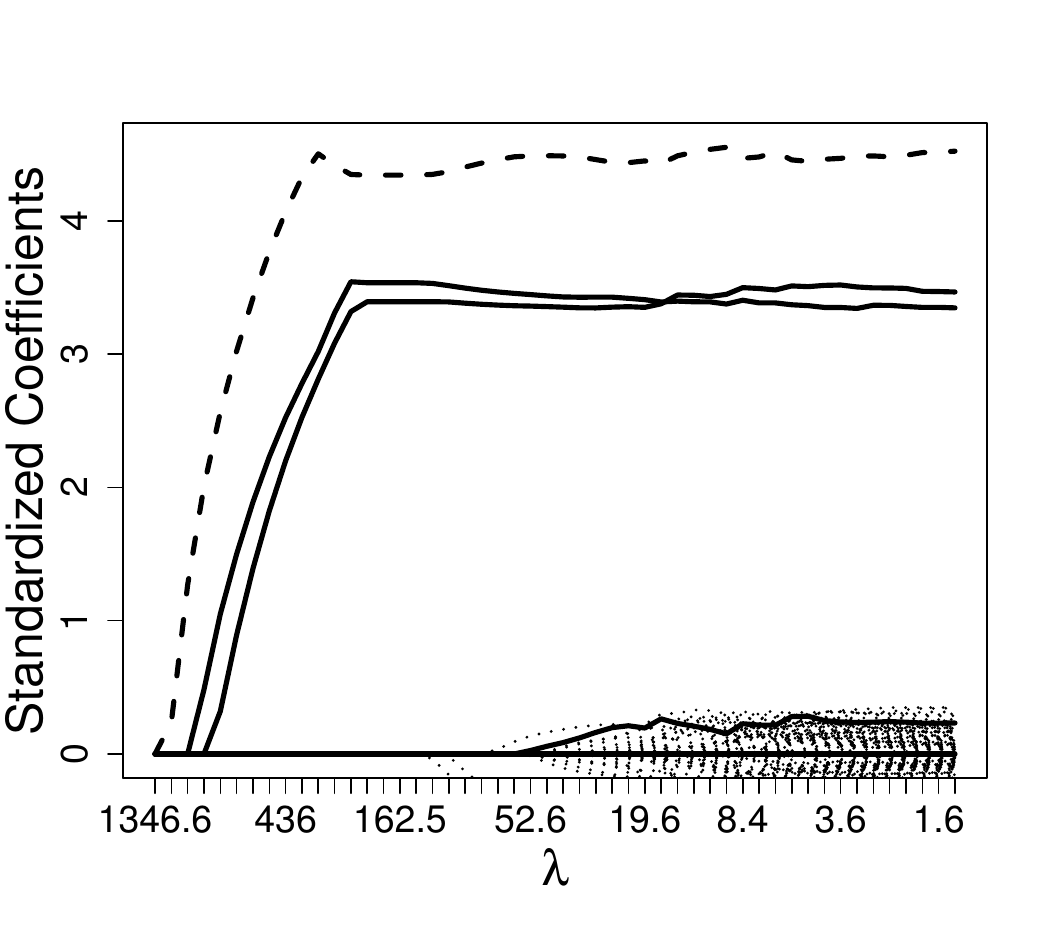}}
\subfigure[OLS post Lasso]{
\includegraphics[width=.23\textwidth,height=.23\columnwidth]
{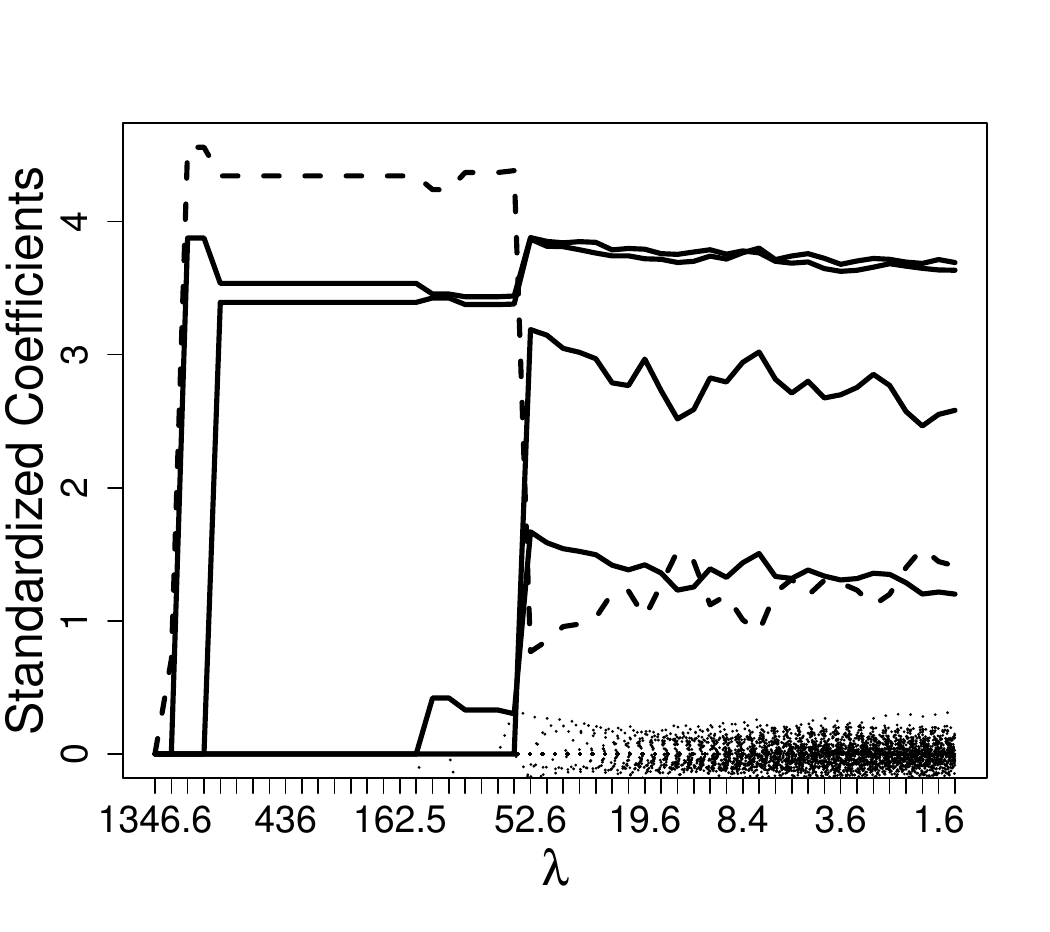}}
\caption{An example to illustrate eight methods' (in) consistency in model selection. The solid lines stand for the relevant variables; the dashed line stands for the variable which is irrelevant but highly correlated with the relevant variables; the dotted lines stand for other irrelevant variables.}
\label{fig:method}
\end{figure}

We use a simple example to demonstrate that many existing methods may not work as well as MSP when irrelevant variables are highly correlated with the relevant variables. Set $(n,p)=(200,400)$ and $\beta$ with 4 nonzero entries. In this example there exists a variable which is irrelevant but highly correlated with the relevant variables, hence the irrepresentable condition fails. Figure~\ref{fig:method} shows eight methods' (in)consistency in model selection: MSP, Lasso \citep{tibshirani1996lasso}, LLA \citep{zou2008one, fan2014strong}, MCP \citep{zhang2010mcp}, SCAD \citep{fan2001variable}, Adaptive Lasso \citep{zou2006adaptive}, OLS post Lasso \citep{belloni2013least} and Capped$-l_1$ \citep{zhangtong2010}. As shown in Figure~\ref{fig:method}, except for MSP, all other methods pick up this irrelevant variable first and never shrink it back to zero. MSP performs similarly when $\lambda$ is large, but when $\lambda$ is small, MSP obtains a stable, accurate estimates and selects the right model. More details of this data example with further comparisons can be found in the simulation studies in Section 4.

\subsection{Relationship to other methods}
There are many widely used methods that estimate regression coefficients for sparse linear models well. In this section, we analyze the MSP solution and describe how our approach differs from these methods, more specifically, how MSP can select the right model when there exist strong correlations between the irrelevant and relevant variables.

Normally, $ \lambda $ is the key to control the amount of regularization, but the proposed method intends to use the iterations to do the controlling rather than using $ \lambda $. Note for any $ y $, $ X $ and $ \lambda $, the solution of the $ k $th iteration of MSP is given by
\[\hat \beta_{(\mA^{[k]})^c} = 0 \ \text{and} \ \hat \beta_{\mA^{[k]}} = (X_{\mA^{[k]}}^\T X_{\mA^{[k]}})^{-1} (X^\T_{\mA^{[k]}} y - \dfrac{\lambda s_{\mA^{[k]}}}{\hat \beta^{[k-1]}_{\mA^{[k]}}}),\]
where $ s_{\mA^{[k]}}$ is the vector of signs of $ \hat \beta_{\mA^{[k]}} $, and the equation on the right may be expressed as
\begin{equation}\label{eq A}
\hat \beta_{\mA^{[k]}} = \hat \beta^\ols_{\mA^{[k]}} - \dfrac{\lambda (X_{\mA^{[k]}}^\T X_{\mA^{[k]}})^{-1} s_{\mA^{[k]}}}{\hat\beta^{[k-1]}_{\mA^{[k]}}},
\end{equation}
where $ \hat \beta^\ols_{\mA^{[k]}} $ is the OLS estimator on the set $ \mA^{[k]} $. For $j=1,\ldots,p$, we make the following notes on the relevant and the irrelevant predictors respectively:
\begin{itemize}
\item Assume $ \beta_j=0 $ and the first step of the MSP algorithm (i.e. Lasso) did not shrink its estimate to zero. Reviewing the estimation properties of the Lasso \citep{meinshausen2009lasso,negahban2012unified}, under the restricted eigenvalue condition and the proper choice of $ \lambda $, this first iteration estimate is bounded by
\[ |\hat \beta^{[1]}_j| = O((\log p/n)^{1/2}) \]
with high probability. At the same time, it is not difficult to verify that $ |\hat \beta^\ols_j| $ has the same bound. According to \eqref{eq A}, given $ \lambda $, the penalty term for the $ j $th variable increases at the rate $ (n/\log p)^{1/2} $ while $ \hat \beta^\ols_j $ is bounded by $ M (\log p/n)^{1/2} $ with some positive constant $ M $. Thus, the associated variable will be deleted from the active set in a finite number of steps, which we have found is typically few.
	\item Assume $ \beta_{j} \neq 0 $ satisfying the Beta-min condition for the nonzero coefficients (see (C.1) in the next section). Following the above argument, there will be a gap between its estimate and 0.  Since $\hat \beta^\ols_{j} $ is bounded away from zero
  and the penalty term will change little after several iterations, the algorithm will stabilize when all the irrelevant variables have been deleted.
\end{itemize}

To explain the difference between our method and others, we consider two examples, MCP \citep{zhang2010mcp} and LLA \citep{zou2008one,fan2014strong} for illustration.
For the MCP, the method essentially uses a large penalty for variables whose estimated coefficients are close to zero and no penalty when the estimated coefficients are large. In the high-dimensional setting, however, by chance there often exist a few irrelevant variables whose estimated coefficients are not close to zero, and this is especially the case when the irrelevant variables are strongly correlated with the relevant variables.  In practice, under such situations, a one-step procedure is often not sufficient to remove all irrelevant variables while keeping all relevant variables.  See Figure~\ref{fig:method}.  LLA, on the other hand, is an iterative method, but in each iteration, the method deals with the entire set of predictors, and since the number of irrelevant variables is always much larger than that of relevant variables, the iteration doesn't really help in terms of choosing an appropriate value of the tuning parameter in comparison with one-step methods, especially when irrelevant variables are strongly correlated with relevant variables.

\section{Theoretical results}
We first define some notation. Without loss of generality, we assume the columns of $X$ are standardized: $X^T\mathds{1}=0$ and $X^T_jX_j/n = 1$ for $j = 1,\ldots,p$.
Let $S \equiv \{ j\in\{1,...,p\}: \beta_j \neq 0 \}$, $ |S| = q $;
let $C=X^TX/n$, $C_{SS}=X^T_SX_S/n$ and $C_{S^cS}=X^T_{S^c}X_S/n$. To state our theoretical results, we need the following assumptions.
\begin{enumerate}
\item[] (C.1) Beta-min condition: assume $1/2<c_3<1$ and there exists a positive constant $ K_1 $ that
\[ \min_{j \in S} |\beta_j| > K_1 n^{(c_3-1)/2}. \]
\item[] (C.2) Restricted Eigenvalue (RE) condition: there exists a positive constant $K_2$ that
\[ v^TCv \geqslant K_2\|v\|^2_2, \]
for all $ v \in G(S) $ where $ G(S) :=\{ v \in \mathds {R}^p: \|v_{S^c}\|_1 \leqslant 3\|v_{S}\|_1 \}$.
\end{enumerate}

(C.1) requires a small gap between $\beta_S$ and $0$. It allows $|\beta_j| \rightarrow 0$ when $n \rightarrow \infty$ but at a rate that can be distinguished. (C.1) has appeared frequently in the literature for proving model selection consistency, e.g. \citet{zhao2006lasso} and \citet{zhang2010mcp}. (C.2) is usually used to bound the $l_2$-error between coefficients and estimates \citep{meinshausen2009lasso,negahban2012unified}, and is also the least restrictive condition of similar types, e.g. the restricted isometry property \citep{candes2007dantzig} and the partial Riesz condition \citep{zhang2008sparsity}. It has been proved that (C.2) holds with high probability for quite general classes of Gaussian matrices for which the predictors may be highly correlated, in which case the irrepresentable condition or the restricted isometry condition may be violated with high probability \citep{raskutti2010restricted}.

We consider the following dimensions, in particular $p =O(\exp(n^{c_1}))$ and $q = O(n^{c_2})$ where $c_2<1/3$ and $0 \leqslant c_1+c_2<c_3$. As a preparatory result, the following proposition shows that the first step of MSP selects an active set $\mA^{[1]}$ containing the true set with high probability.
\begin{prop}\label{prop}
Suppose (C.1) and (C.2) hold. Set $\lambda_0 = 4\sigma (n \log p)^{1/2}$. Considering the first step of MSP, $\hat \beta^{[1]}(\lambda_0)$ and the corresponding set $\mA^{[1]}$, we have
\begin{equation}\label{eq prop1}
	P(S \subseteq  \mA^{[1]}) \geqslant  1-1/p.
\end{equation}
\end{prop}

\rmk{Since the first step estimator of MSP is the Lasso, Proposition~\ref{prop} can be seen as proving a property for the Lasso estimator. We obtain this result by using the bound of $l_2$-norm error between $\beta$ and $\hat \beta^{[1]}$, which is known from past work, e.g. \citet{meinshausen2009lasso} and \citet{negahban2012unified}. Proposition~\ref{prop} supports the backward deletion strategy of MSP, which removes the variables that do not belong to $\mA^{[1]}$ as they are irrelevant with high probability.}

\rmk{We set $\lambda_0 = 4\sigma (n\log p)^{1/2}$, which is the same as that for Lasso in order to achieve the error bound, while Lasso's model selection consistency requires a larger tuning parameter, i.e. $K n^{(1+c_4)/2}$ where $c_1 < c_4$.  Hence when $n$ is large, the estimation accuracy and selection consistency cannot hold at the same time for Lasso. We solve this problem using an iterative strategy.}

Now we show results on the error bound and the sign consistency of MSP.

\begin{thm}\label{thm}
Suppose (C.1) and (C.2) hold. Set
$\lambda = 4\sigma (n\log n)^{1/2}$. For $(1+3c_2)/2<c_3<1$, with probability at least $1-1/n$, the following error bounds for the estimate $\hat \beta$ hold,
\begin{equation}\label{eq thm}
\begin{tabular}{lr}
	$\|\hat\beta - \beta\|_2 \leqslant \dfrac{8\sigma}{K_2 \cdot K_3}
	\Big(\dfrac{q \log n}{n^{c_3}}\Big)^{1/2}$,\\
	$\|\hat \beta - \beta\|_1 \leqslant \dfrac{32\sigma \cdot q}{K_2 \cdot K_3}\Big(\dfrac{\log n}{n^{c_3}}\Big)^{1/2}$,
\end{tabular}
\end{equation}
where $K_3<K_1$, $K_1$ and $K_2$ are defined in (C.1) and (C.2) respectively. Further, we have:
\[P(sign(\hat \beta) = sign(\beta)) \geqslant 1 - 1/n.\]
\end{thm}

\rmk{Note that $\lambda$ and $\lambda_0$ have different orders under the assumption that \\$p = O(\exp(n^{c_1}))$. If we consider another high dimensional setting, where $p = O(n)$, by setting $\lambda_0 = \lambda = 4\sigma (n\log n)^{1/2}$, we would have the same result as in Theorem~\ref{thm}.
For simplicity, we use the same value for $\lambda_0$ and $\lambda$ in simulation studies and empirical analysis.}

\rmk{The error bound of MSP in \eqref{eq thm} is influenced by the adaptive penalty. We allow $\beta_j$ to converge to 0 in (C.1), i.e., the lower bound of $\beta_j$ is $n^{(c_3-1)/2}$ for $j \in S$.  As a consequence, $n^{c_3/2}$ dominates the denominator of the error bound of MSP rather than $n^{1/2}$. When $c_3$ is close to $1$, the $l_2$-norm error bound is close to the rate $(q \log n /n)^{1/2}$.}

Now suppose (C.1) is replaced by the following condition:
\begin{enumerate}
\item[] (C.1)* For the nonzero coefficients, let $c = \min_{j \in S} |\beta_j|$ and assume $1/c<\infty$.
\end{enumerate}

Note (C.1)* sets a lower bound for $\beta_S$ where $c$ is allowed to be any positive constant. This condition has also appeared frequently in the literature, e.g. \cite{huang2008adaptive}. Replacing (C.1) by (C.1)* and considering the following dimensions: $p = O(\exp(n^{c_1}))$ and $q= O(n^{c_2})$ where $0<c_1+c_2<1$ and $0<c_1<1/3$, Corollary~1 shows that the $l_1$-error and the $l_2$-error of the MSP estimator achieve the rate $q(\log n/n)^{1/2}$ and $(q \log n /n)^{1/2}$, respectively.

\begin{coro}\label{coro}
Suppose (C.1)* and (C.2) hold. Set
$\lambda = 4\sigma (n\log n)^{1/2}$. With probability $1-1/n$, the following error bounds hold for $\hat \beta$:
\begin{equation}\label{eq coro}
\begin{tabular}{lr}
$\|\hat\beta - \beta\|_2 \leqslant \dfrac{8\sigma}{c K_2}\Big(\dfrac{q\log n}{n}\Big)^{1/2}$,\\
$\|\hat \beta - \beta\|_1 \leqslant \dfrac{32\sigma \cdot q}{c K_2}\Big(\dfrac{\log n}{n}\Big)^{1/2}$.
\end{tabular}
\end{equation}
Further, we have:
\[ P(sign(\hat \beta) = sign(\beta)) \geqslant 1 - 1/n. \]
\end{coro}

\rmk{Note the Gaussian assumption on the error term in the linear regression model can be relaxed by a subgaussian assumption. Specifically, there exist constants $K$, $k>0$ such that for $i=1, \ldots, n$,
\[ P(|\epsilon_i| \geqslant t)\leqslant Ke^{-kt^2}, \
\forall t \geqslant 0. \]}

\section{Simulation studies}

In this section, we use simulation studies to demonstrate the performance of the proposed method: 1) the first part illustrates MSP's consistency in model selection; 2) the second part compares the performance of the proposed method with those of several existing methods, and also analyzes the stability of MSP with respect to the tuning parameter $\lambda$; and 3) the third part evaluates the computational time of different methods.

Of the existing methods that are compared with the proposed method, we choose three one-step methods, Lasso \citep{tibshirani1996lasso}, SCAD \citep{fan2004nonconcave} and MCP \citep{zhang2010mcp}, and four multi(two)-step methods, Adaptive Lasso \citep{zou2006adaptive} (denoted as Alasso), OLS post Lasso (denoted as Plasso) \citep{belloni2013least}, Capped$-l_1$ \citep{zhangtong2010} and LLA \citep{zou2008one, fan2014strong}. In addition, we also compare with a Bayesian method, SSL \citep{rovckova2018spike}.  We use the R package SSLASSO to run SSL; results of MCP, SCAD and LLA are obtained using the R ncvreg package \citep{breheny2011coordinate}, and results of other methods are based on the R glmnet package \citep{friedman2010regularization}.

We consider the following linear regression model for simulation studies
\[y_i = \sum_{j = 1}^{p} x_{ij} \beta_{j} + \epsilon_i, \ i = 1,\ldots,n,\]
where $x_{ij}$ are generated from a multivariate normal distribution $N (0, \Sigma)$ and $\epsilon_i$ is generated from $ N(0, 1) $.  Four regression coefficients are set as nonzero, specifically $(\beta_2, \beta_3, \beta_4, \beta_p) = (2,4,4,4)$, and others are set to zero. We consider the case where some irrelevant variable is highly correlated with the relevant variables. Specifically, we set
\[ x_{i1} = \dfrac{7}{8} x_{ip} + \dfrac{3}{8} x_{i2} + \dfrac{1}{8} x_{i3} + \dfrac{1}{8} x_{i4} + \dfrac{1}{8} x_{i5} + \dfrac{1}{8} x_{i6} + \dfrac{1}{8} x_{i7} + \dfrac{1}{8}e_i, \]
where $e_i$ is generated from $ N(0,1)$.  Denote the covariance matrix of the last $(p-1)$ variables as $\Sigma_{-1}$. We consider two scenarios: (1) $\Sigma_{-1} = \text{I}$, and (2) $\Sigma_{jj'} = 0.5^{|j-j'|}$, where $j = 2,\ldots,p$.

In both scenarios, the RE condition (C.2) holds while the irrepresentable condition fails. Recall the irrepresentable condition states that: there exists a positive constant $\eta >0$ such that
\[ \|C_{S^cS}C^{-1}_{SS}\text{sign}(\beta_S)\|_{\infty} \leqslant 1-\eta. \]
With $X_S = (X_2,X_3,X_4,X_p)$ and $X_{S^c} = (X_1,X_5,...,X_{p-1})$, it is not difficult to check that the irrepresentable condition does not hold. Figure~\ref{fig:method} in Section~\ref{method} shows the results from one typical simulation repetition when $(n,p)=(200,400)$ under scenario 1.

We compare both the estimation and selection performances of the nine methods mentioned above.
The $l_2$-norm ($\|\hat \beta - \beta\|_2$) and the $l_1$-norm errors ($\|\hat \beta - \beta\|_1$) are computed. We also report the estimated number of nonzero coefficients (NZ), as well as the false positive rate (FPR) and the true positive rate (TPR), which are respectively defined as
\begin{eqnarray*}
\mbox{FPR} &=& \dfrac{|j \in \{1,...,p\}: \hat \beta_j \neq 0 \  \text{and} \ \beta_j = 0|}{|j \in \{1,...,p\}:  \beta_j = 0|}, \\
\mbox{TPR} &=& \dfrac{|j \in \{1,...,p\}: \hat \beta_j \neq 0 \  \text{and} \ \beta_j \neq 0|}{|j \in \{1,...,p\}:  \beta_j \neq 0|}.
\end{eqnarray*}

\subsection{Model selection}
\begin{figure}[!htp]
\centering
\subfigure[MSP, $p = 40$]{
\includegraphics[width=.32\textwidth,height=.23\columnwidth]
{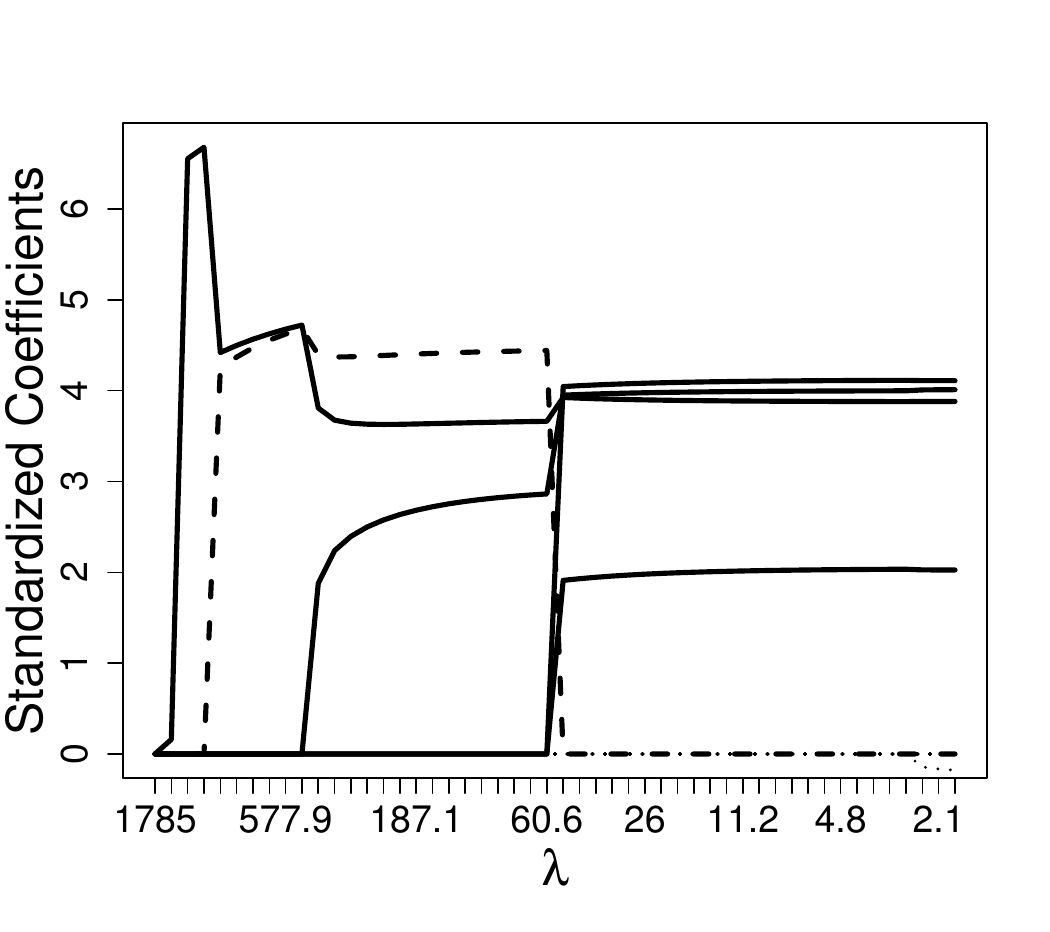}}
\subfigure[MSP, $p = 400$]{
\includegraphics[width=.32\textwidth,height=.23\columnwidth]
{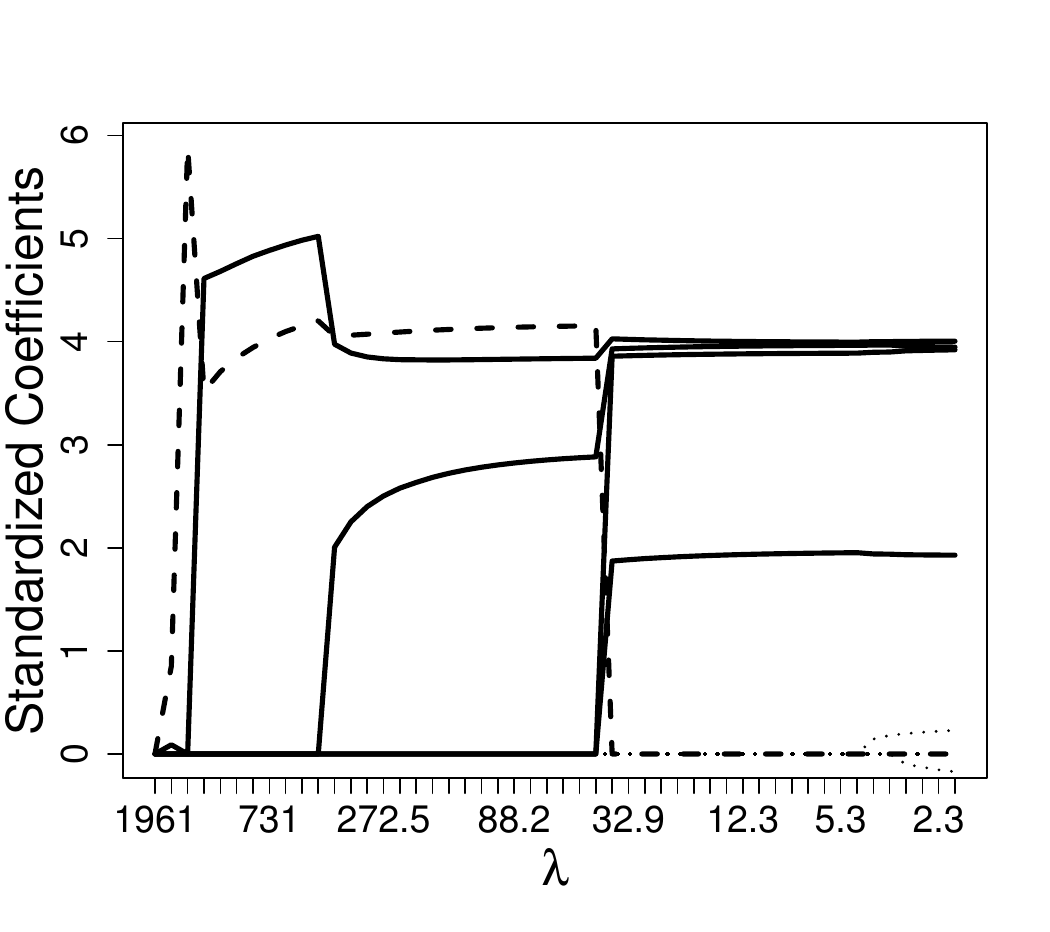}}
\subfigure[MSP, $p = 4000$]{
\includegraphics[width=.32\textwidth,height=.23\columnwidth]
{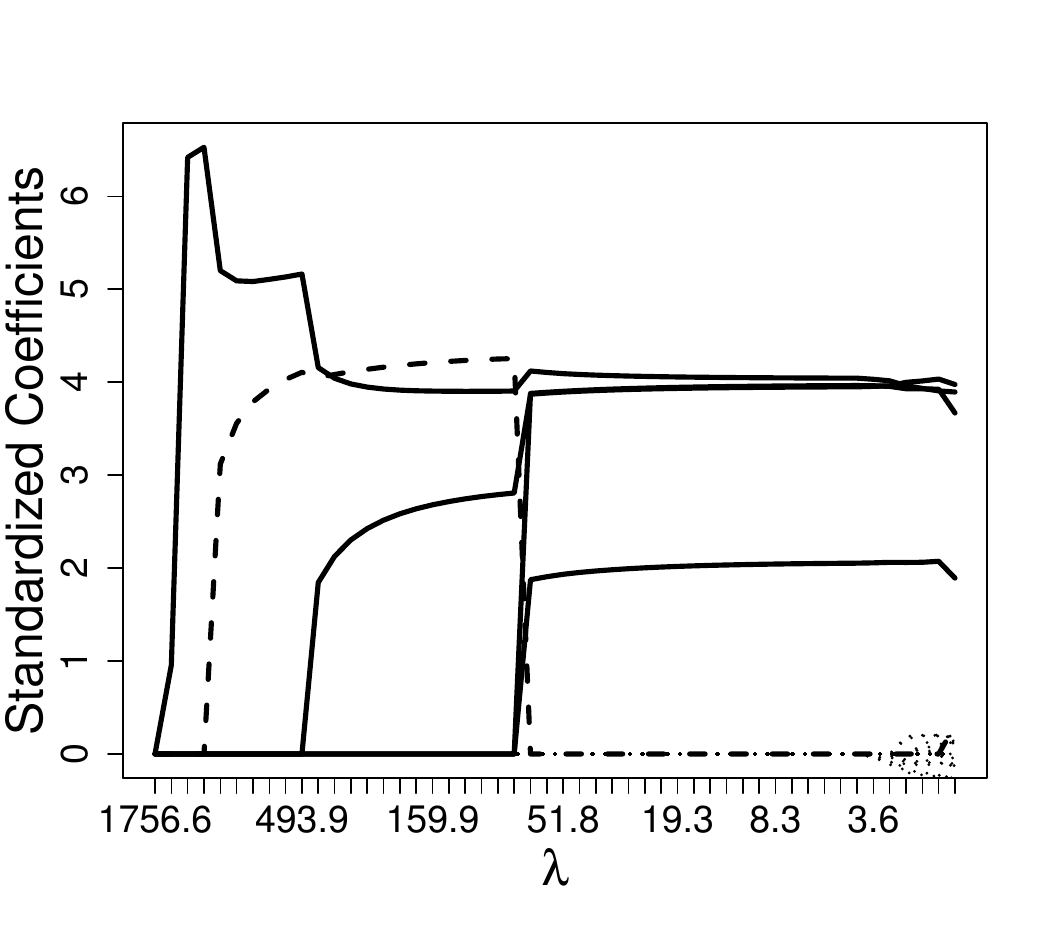}}
\subfigure[Lasso, $p = 40$]{
\includegraphics[width=.32\textwidth,height=.23\columnwidth]
{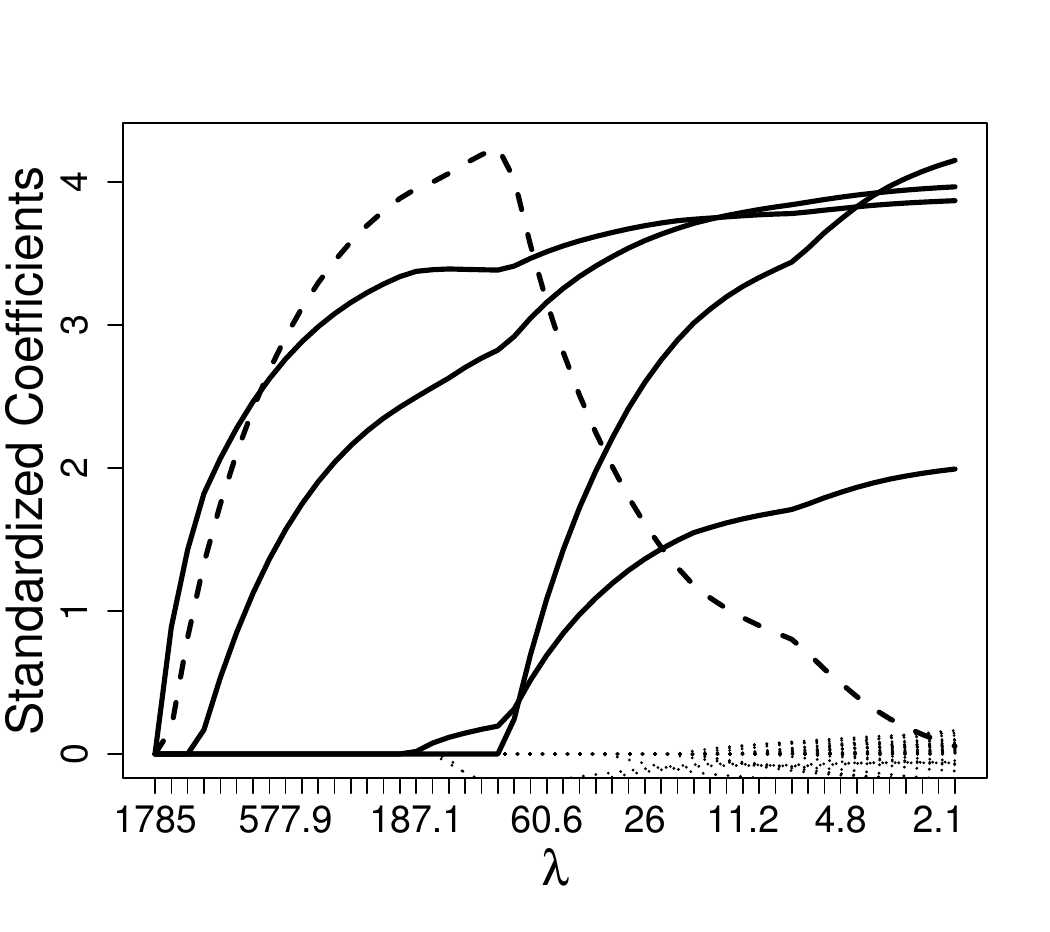}}
\subfigure[Lasso, $p = 400$]{
\includegraphics[width=.32\textwidth,height=.23\columnwidth]
{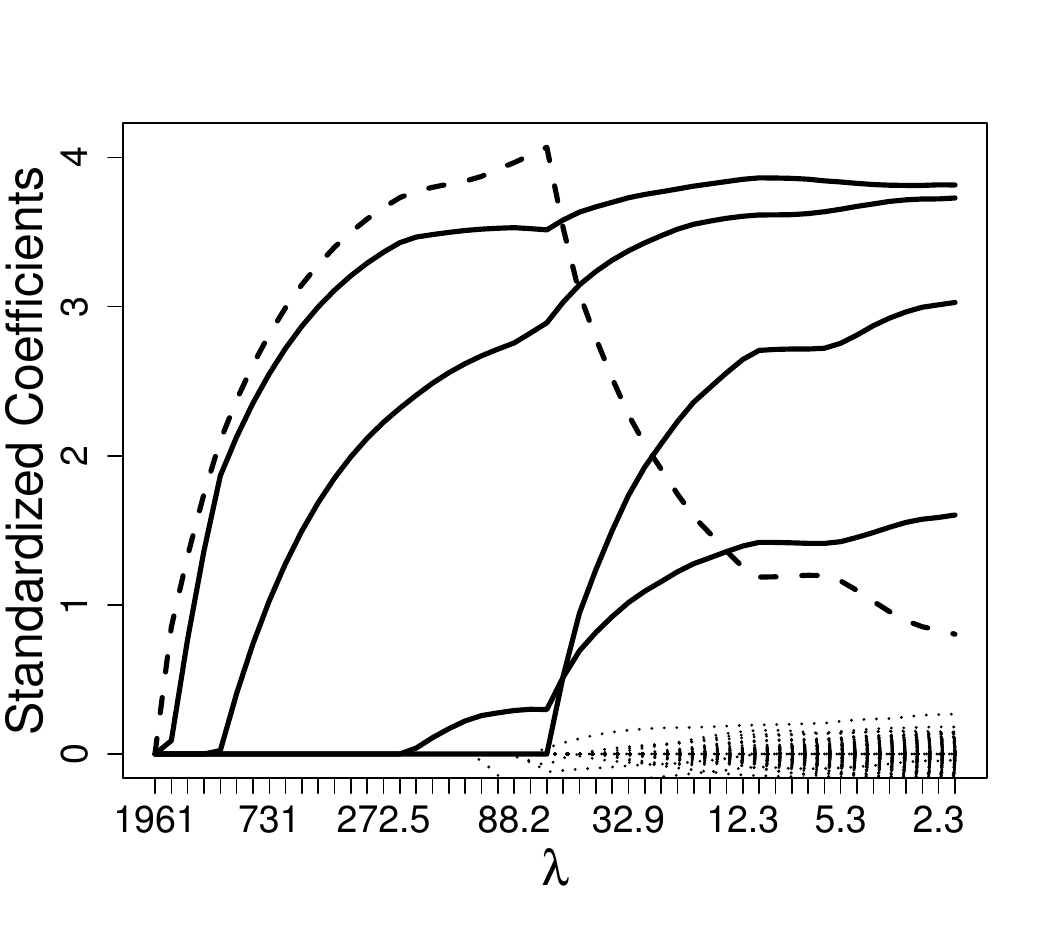}}
\subfigure[Lasso, $p = 4000$]{
\includegraphics[width=.32\textwidth,height=.23\columnwidth]
{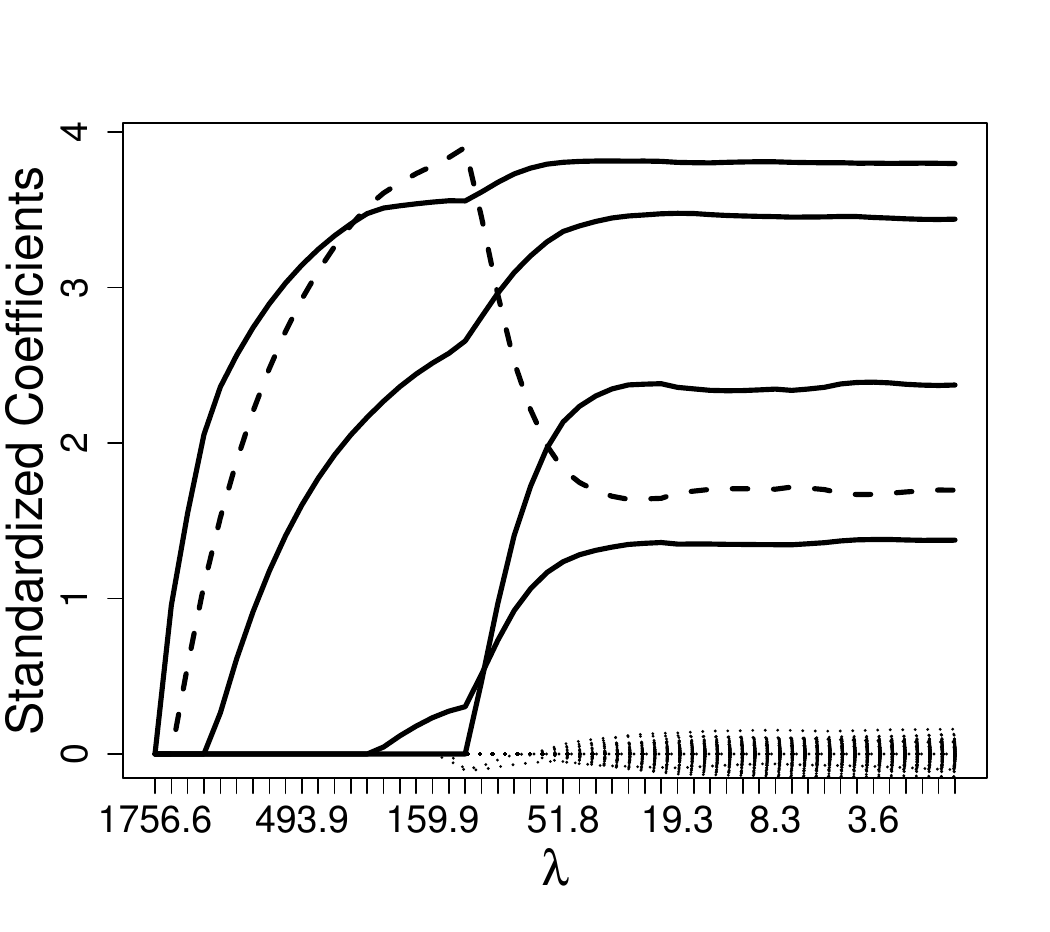}}
\subfigure[PLasso, $p = 40$]{
\includegraphics[width=.32\textwidth,height=.23\columnwidth]
{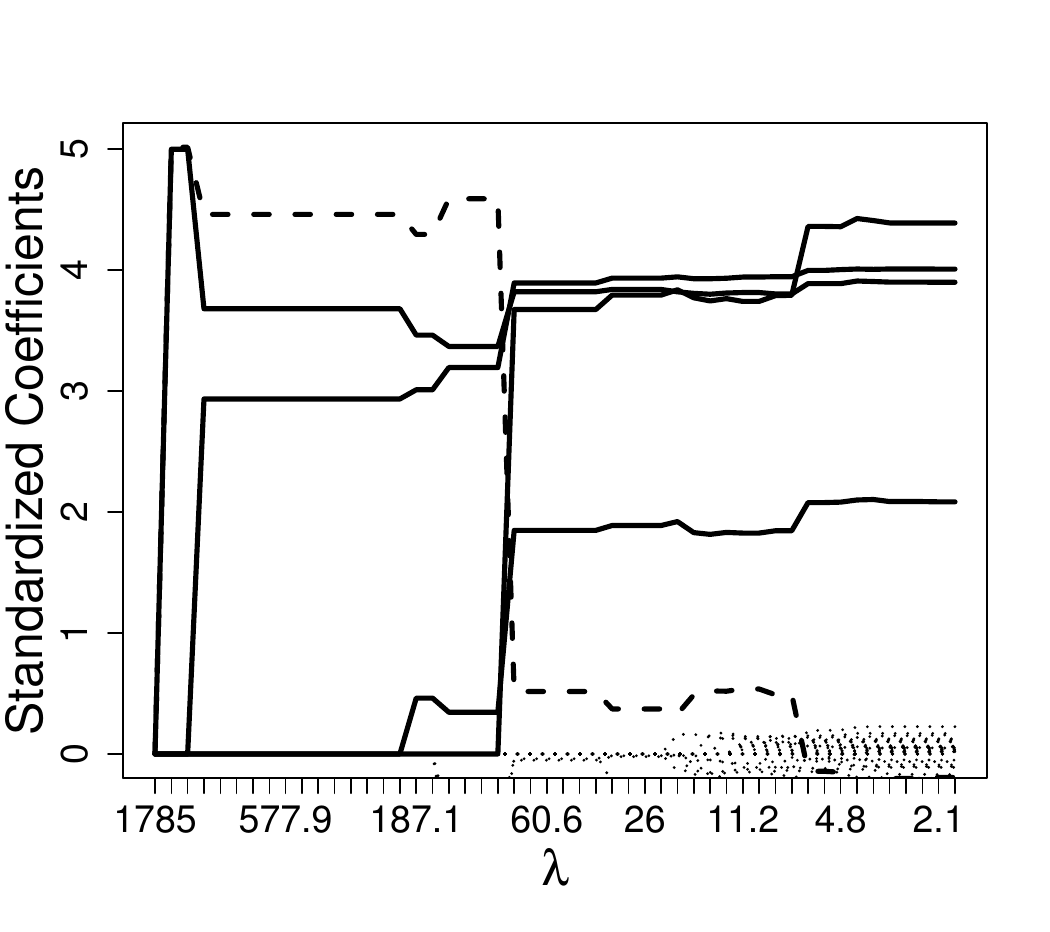}}
\subfigure[PLasso, $p = 400$]{
\includegraphics[width=.32\textwidth,height=.23\columnwidth]
{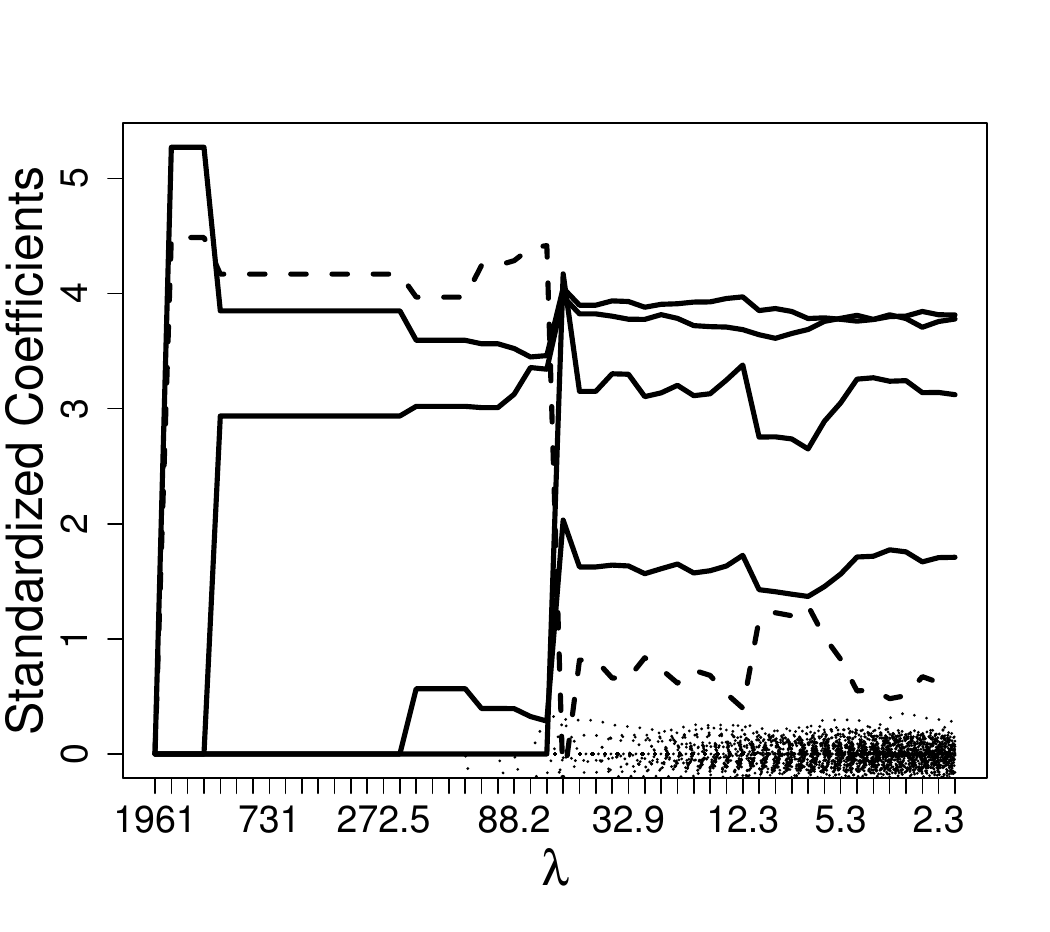}}
\subfigure[PLasso, $p = 4000$]{
\includegraphics[width=.32\textwidth,height=.23\columnwidth]
{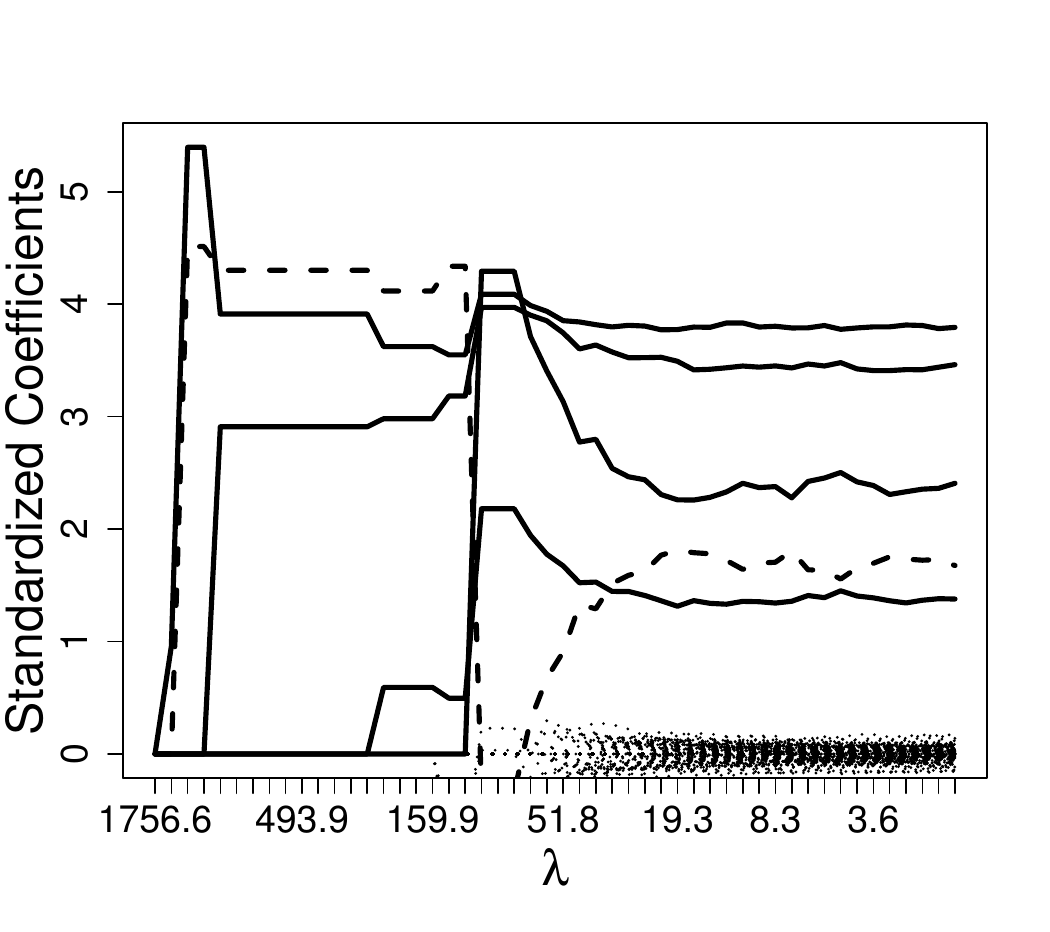}}
\subfigure[SCAD, $p = 40$]{
\includegraphics[width=.315\textwidth,height=.23\columnwidth]
{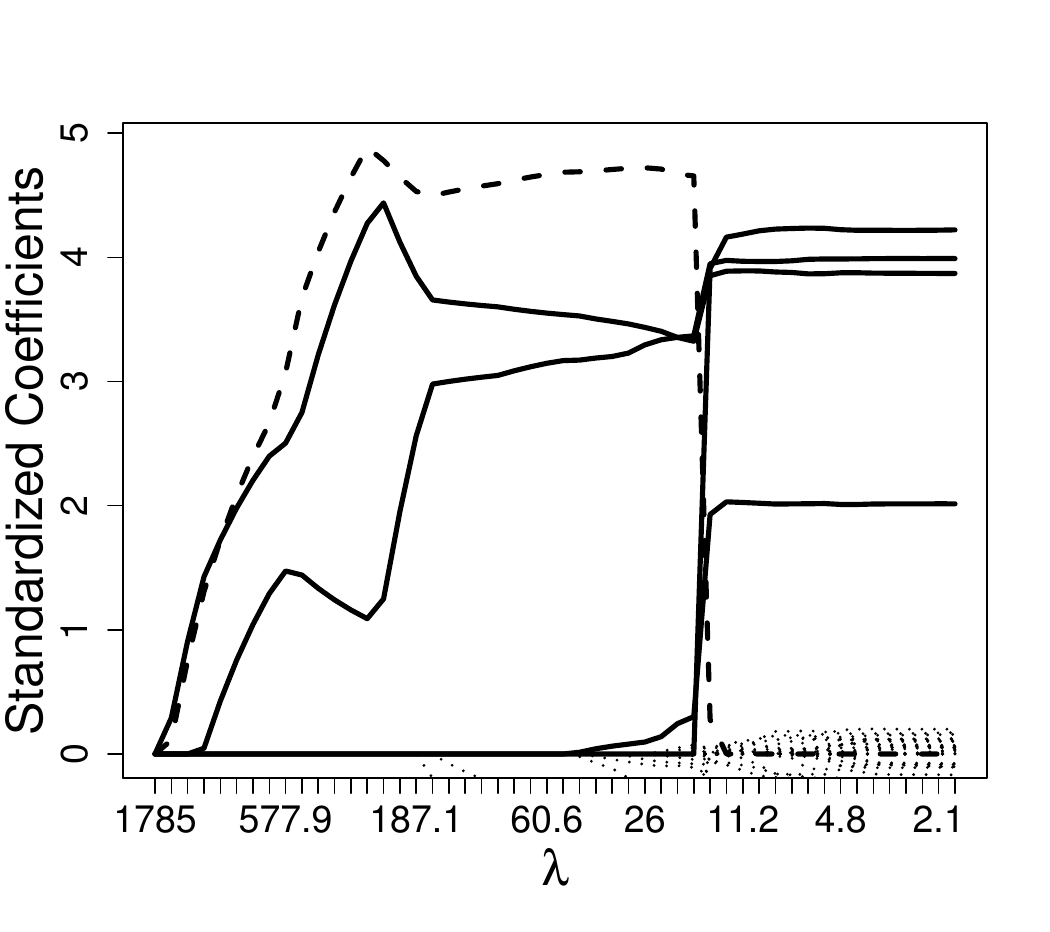}}s
\subfigure[SCAD, $p = 400$]{
\includegraphics[width=.315\textwidth,height=.23\columnwidth]
{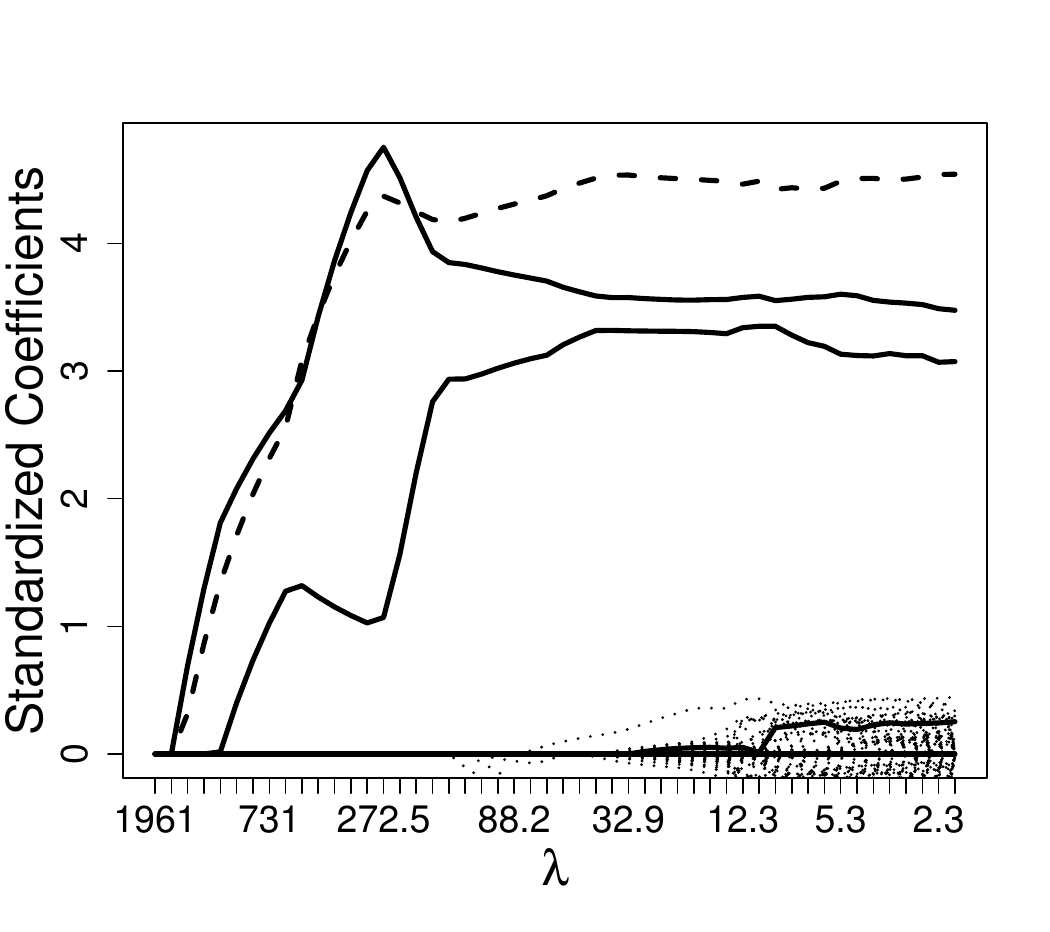}}
\subfigure[SCAD, $p = 4000$]{
\includegraphics[width=.315\textwidth,height=.23\columnwidth]
{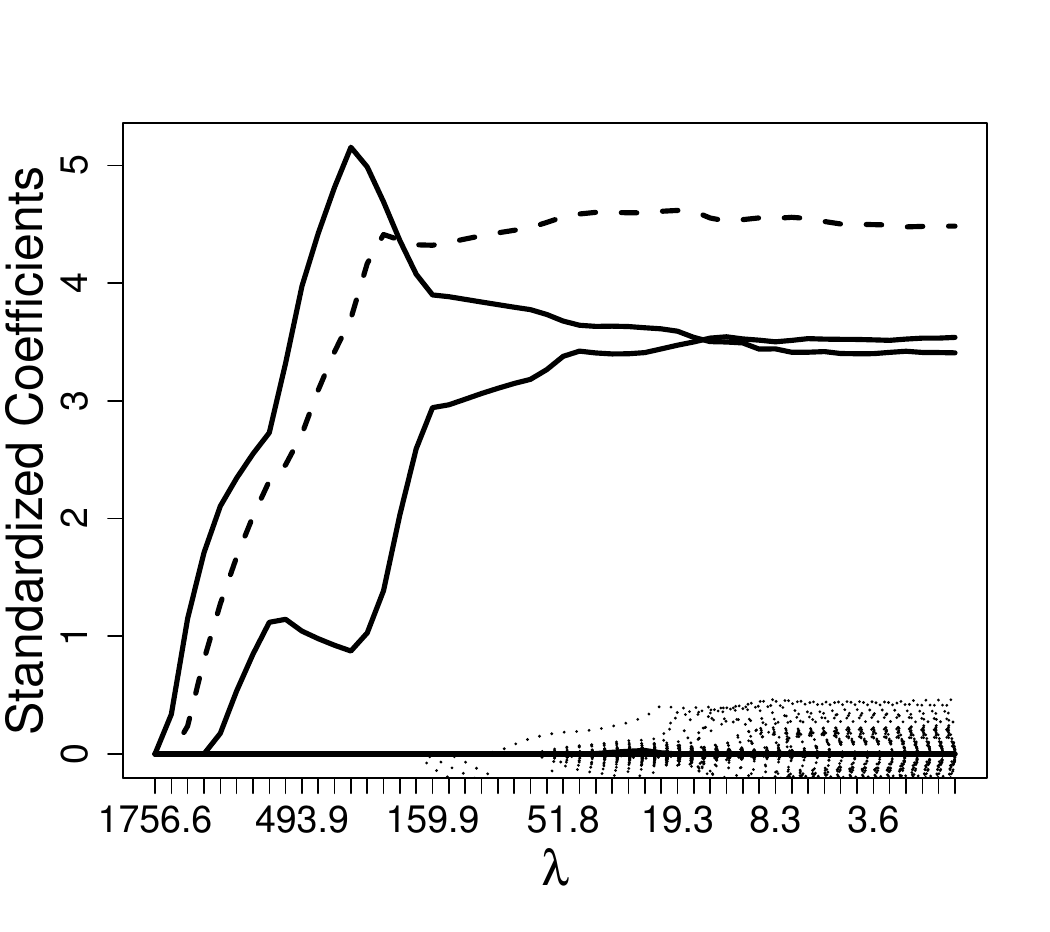}}
\caption{Results for Scenario 2 under three different dimensions.
The dashed line corresponds to $X_1$, which is irrelevant; the dotted lines correspond to other irrelevant variables; the solid lines correspond to the relevant variables.}
\label{fig:2}
\end{figure}

We consider three different dimensions, $p$=40, 400 and 4000, and $n$ is fixed to be 200. Due to lack of space, we only show representative results in the main manuscript and delay other results in the supplementary material. Specifically, Figure~\ref{fig:1} shows the comparison between the proposed MSP, Capped-$l_1$, LLA, MCP and SSL under Scenario 1, while Figure~\ref{fig:2} shows the comparison between MSP and Lasso, PLasso and SCAD under Scenario 2. As we can see in Figure~\ref{fig:1}, when $\lambda$ is large, the first 4 methods select the irrelevant variable $X_1$ (as it is highly correlated with the relevant variables and the response).  When $\lambda$ decreases, in the case of relatively low dimension (left column), LLA and MCP are able to shrink the estimated coefficient for $X_1$ to zero but at the same time select many other irrelevant variables; while in the case of relatively high dimension (middle and right columns), LLA and MCP are not even able to shrink the estimated coefficient for $X_1$ to zero. Capped-$l_1$ performs slightly better as it is able to shrink the estimated coefficient for $X_1$ to zero in both low and high dimensional settings when $\lambda$ is very small but at the same time also selects many irrelevant variables.  SSL chooses the correct model when $p$=40 and 400, however, when $p$ becomes larger, SSL always selects $X_1$ as an important variable.  As a comparison, MSP chooses the exact correct model over a wide range of small values of $\lambda$ in all settings.

The results in Figure~\ref{fig:2} are similar as in Scenario 1: Lasso, PLasso and SCAD are able to shrink the estimated coefficient for $X_1$ to zero when $\lambda$ is small and the dimension is relatively low, but at the same time select many other irrelevant variables, and completely fail to shrink the estimated coefficient for $X_1$ to zero when the dimension is relatively high. The proposed MSP is again able to identify the correct model when $\lambda$ is relatively small in all three considered dimensional settings.

We also want to note that, as one can see in both Figures~\ref{fig:1} and \ref{fig:2}, the MSP solution is quite stable over a wide range of small values of $\lambda$. This implies that MSP requires little tuning, which is a convenient and useful property in practice and different from many other regularization methods that require careful selection of the tuning parameter.

\begin{figure}
\centering
\subfigure[MSP, $p=40$]{
\includegraphics[width=.32\textwidth,height=.21\columnwidth]
{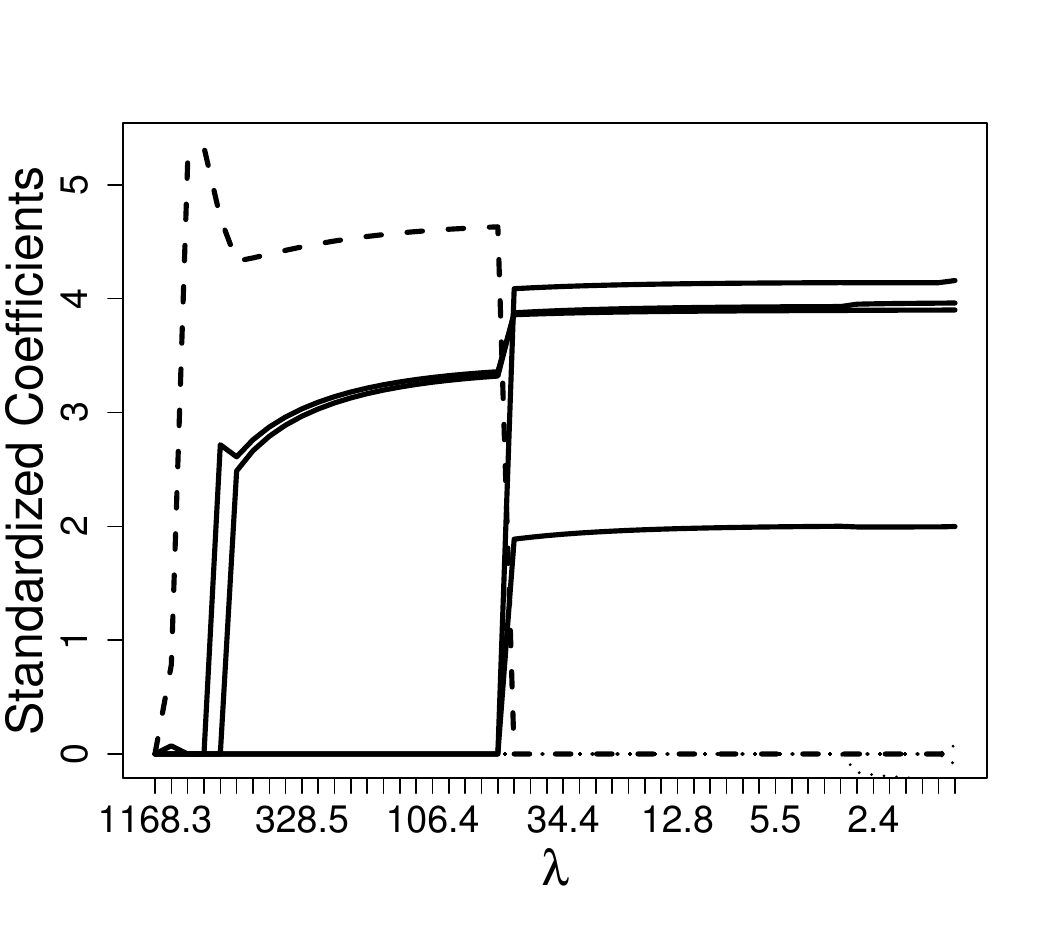}}
\subfigure[MSP, $p=400$]{
\includegraphics[width=.32\textwidth,height=.21\columnwidth]
{./fig/msp400.pdf}}
\subfigure[MSP, $p=4000$]{
\includegraphics[width=.32\textwidth,height=.21\columnwidth]
{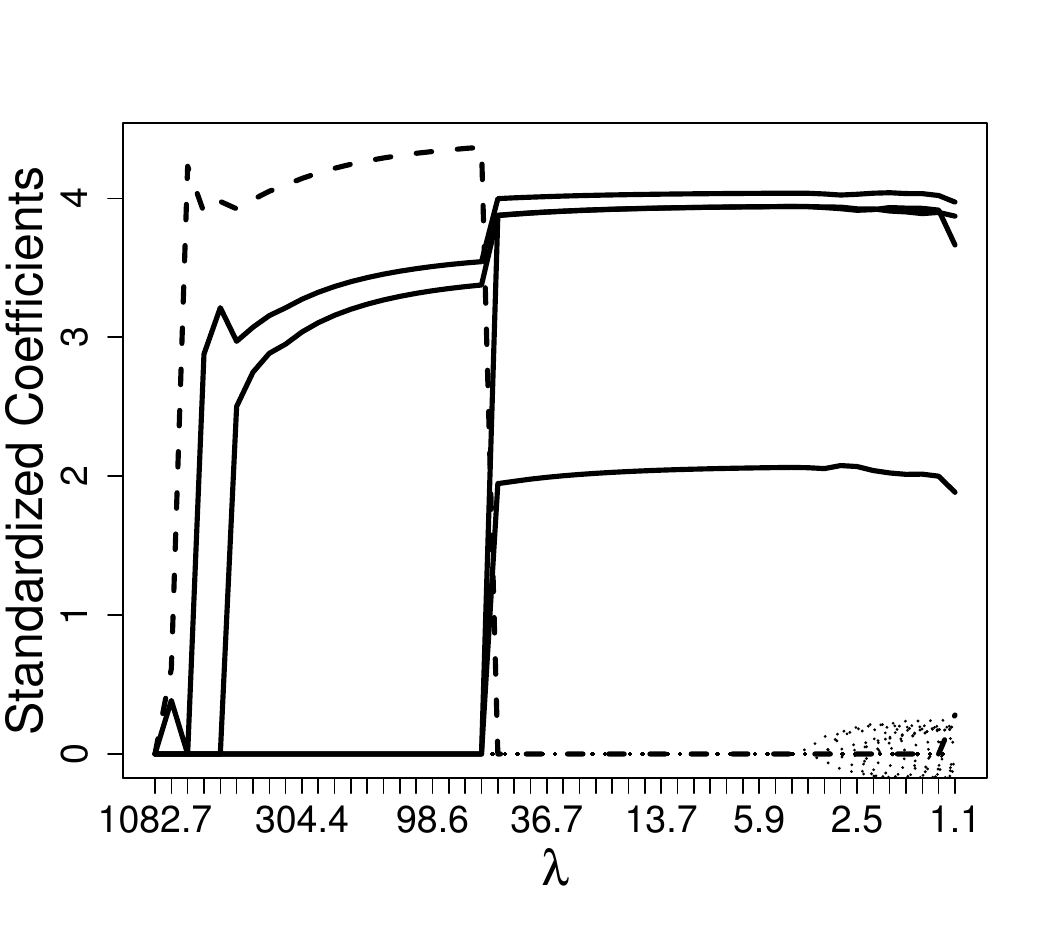}}
\subfigure[Capped-$l_1$, $p=40$]{
\includegraphics[width=.32\textwidth,height=.21\columnwidth]
{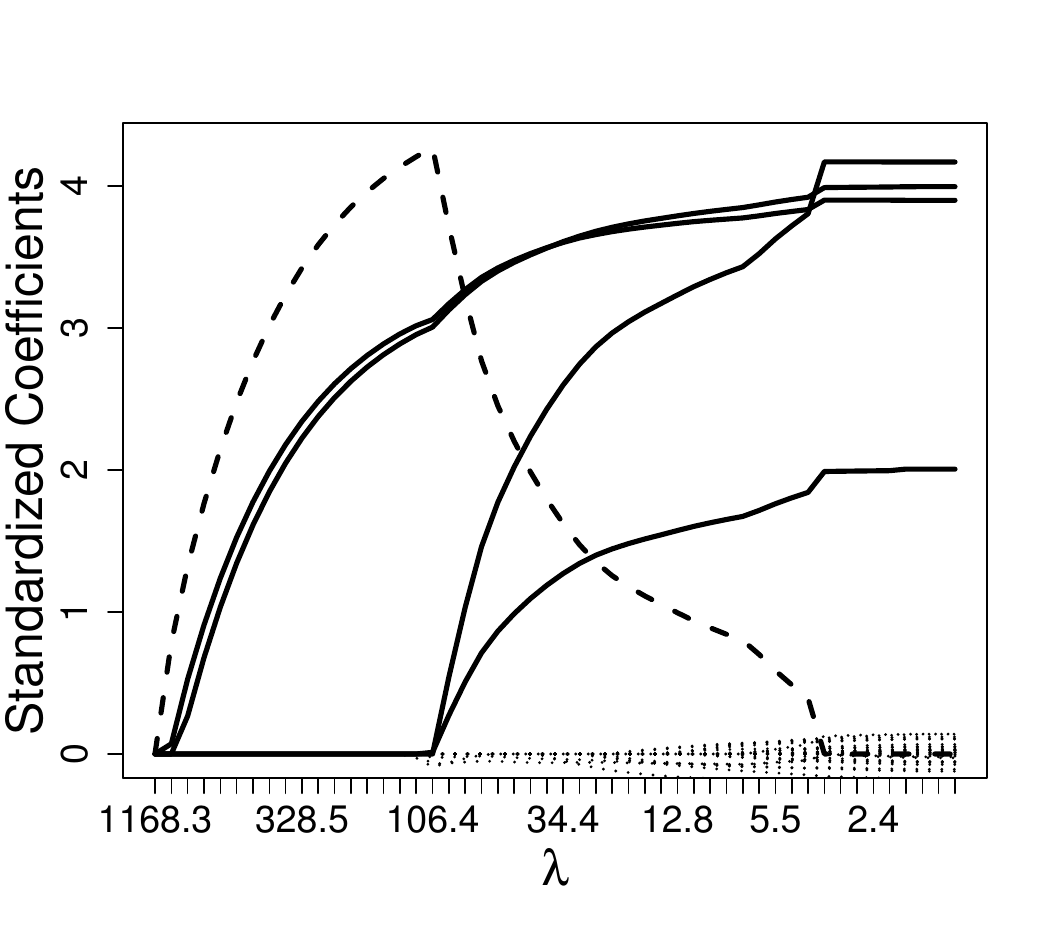}}
\subfigure[Capped-$l_1$, $p=400$]{
\includegraphics[width=.32\textwidth,height=.21\columnwidth]
{./fig/cap400.pdf}}
\subfigure[Capped-$l_1$, $p=4000$]{
\includegraphics[width=.32\textwidth,height=.21\columnwidth]
{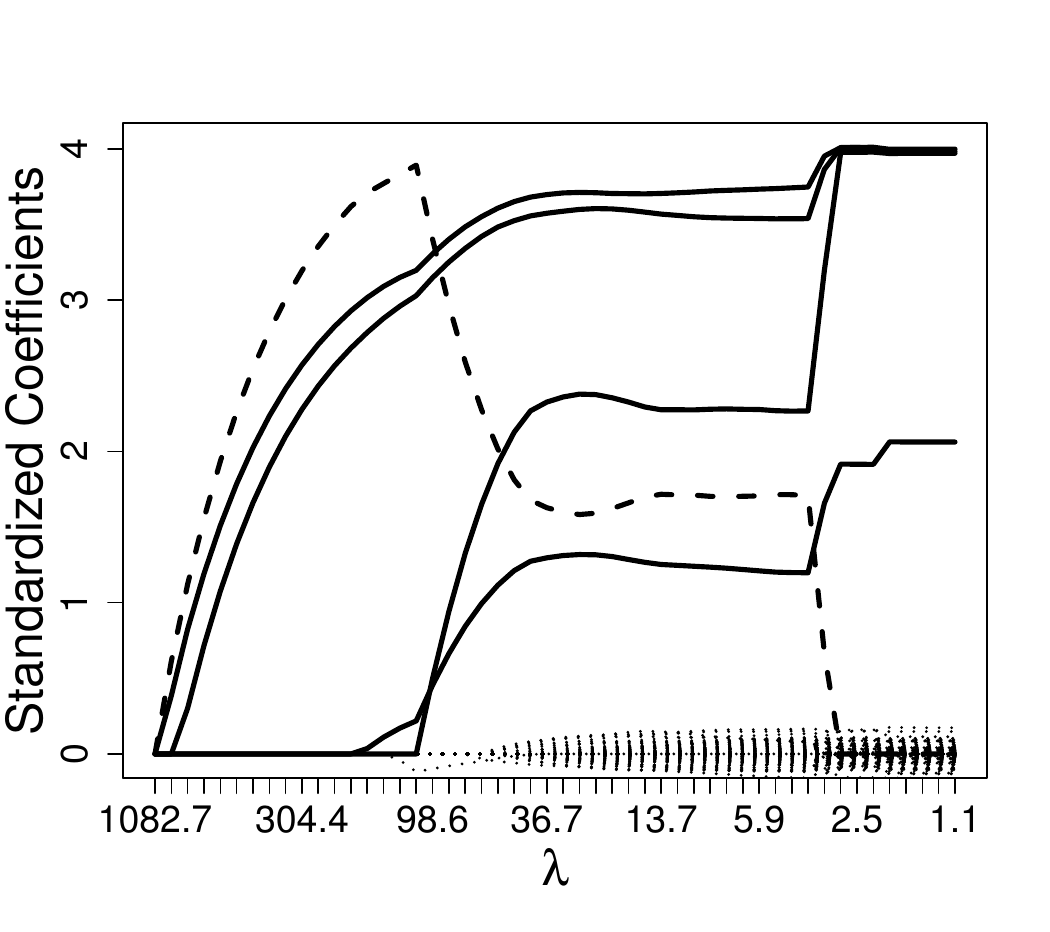}}
\subfigure[LLA, $p=40$ ]{
\includegraphics[width=.32\textwidth,height=.21\columnwidth]
{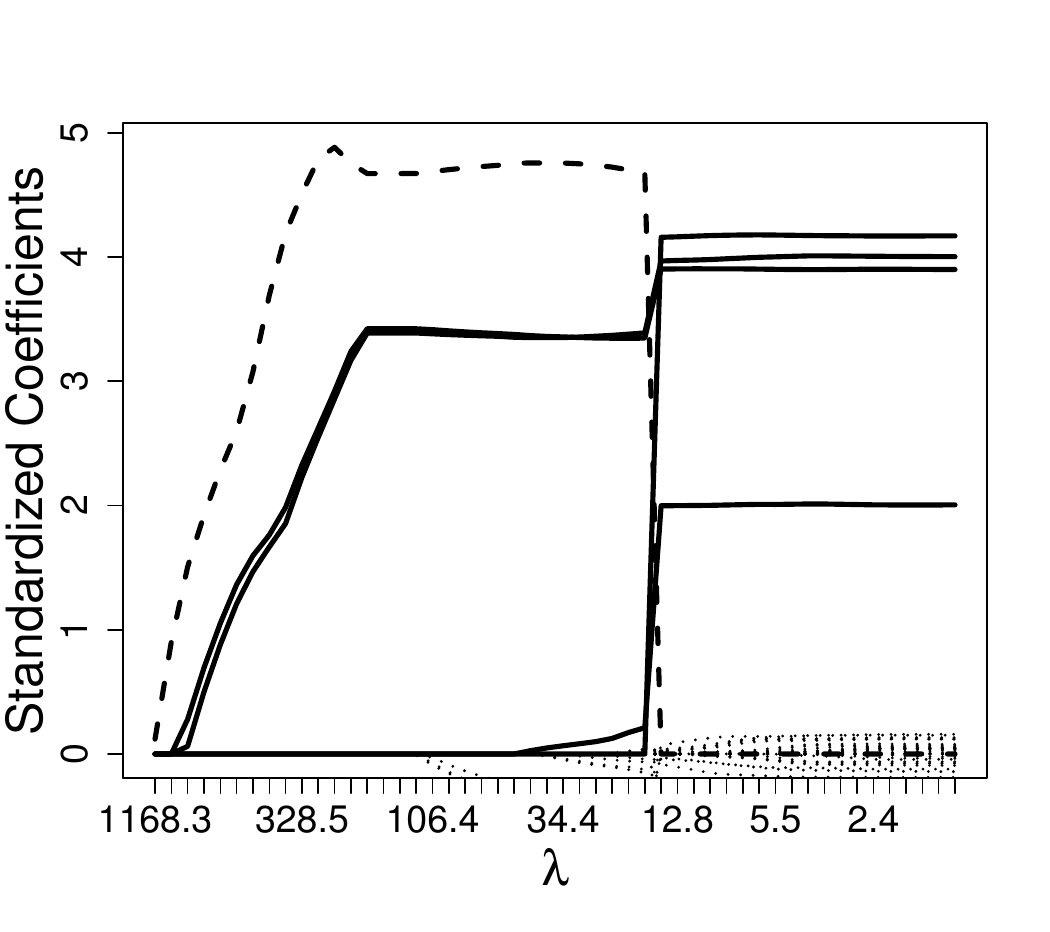}}
\subfigure[LLA, $p=400$]{
\includegraphics[width=.32\textwidth,height=.21\columnwidth]
{./fig/lla400.pdf}}
\subfigure[LLA, $p=4000$]{
\includegraphics[width=.32\textwidth,height=.21\columnwidth]
{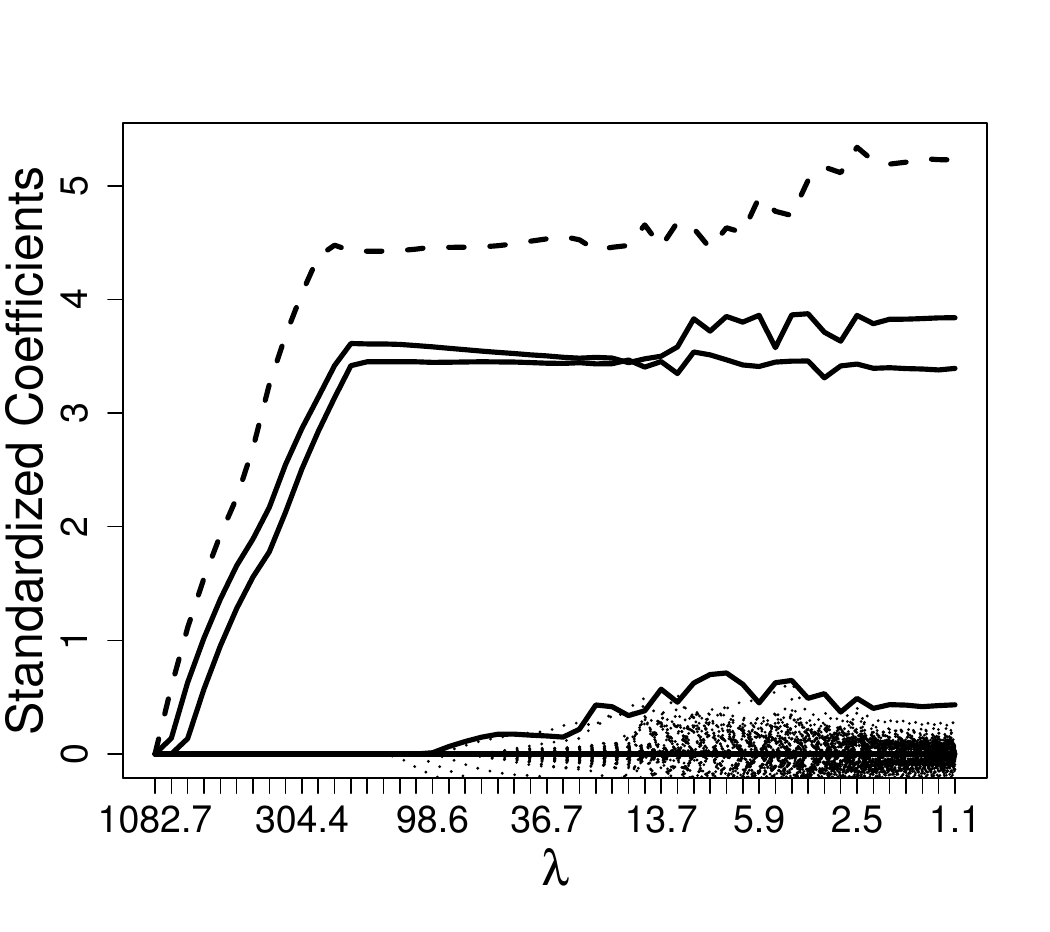}}
\subfigure[MCP, $p=40$]{
\includegraphics[width=.32\textwidth,height=.21\columnwidth]
{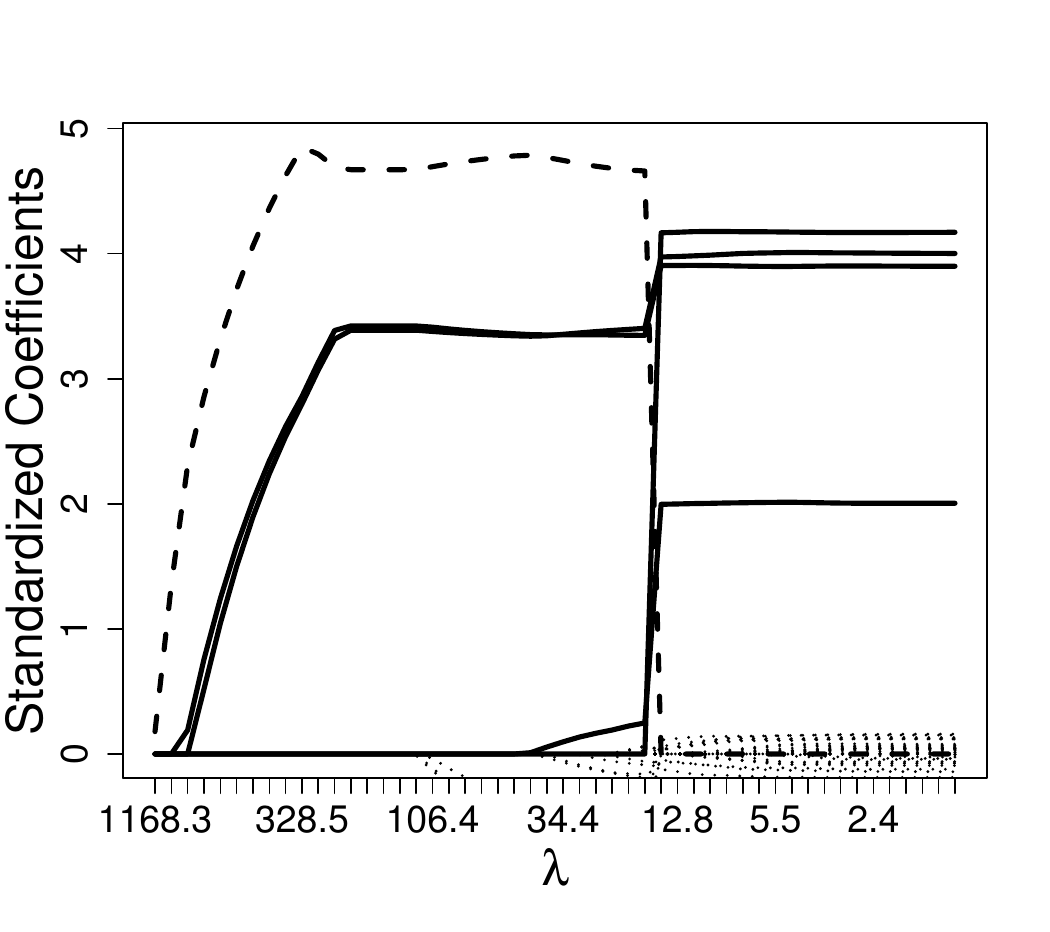}}
\subfigure[MCP, $p=400$]{
\includegraphics[width=.32\textwidth,height=.20\columnwidth]
{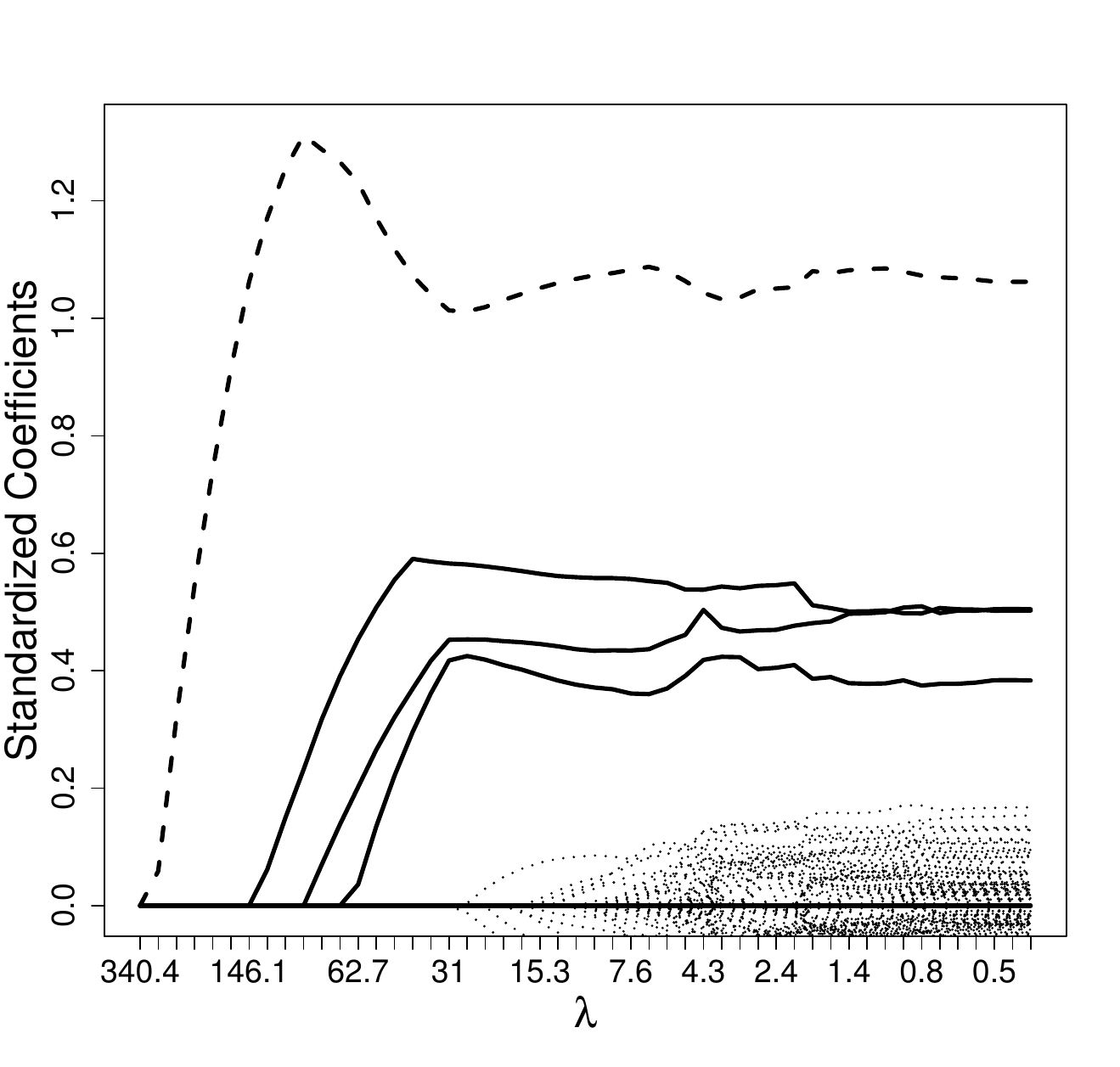}}
\subfigure[MCP, $p = 4000$]{
\includegraphics[width=.32\textwidth,height=.21\columnwidth]
{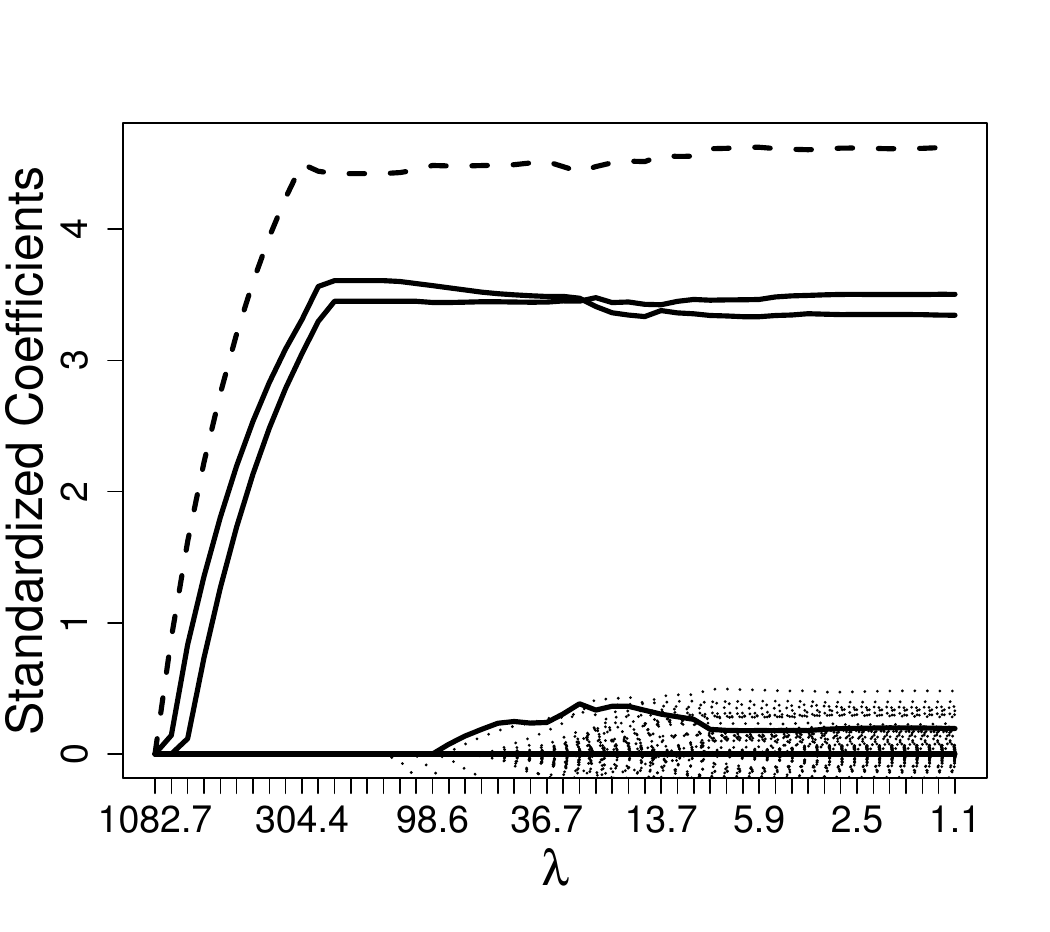}}
\subfigure[SSL, $p=40$]{
\includegraphics[width=.32\textwidth,height=.21\columnwidth]
{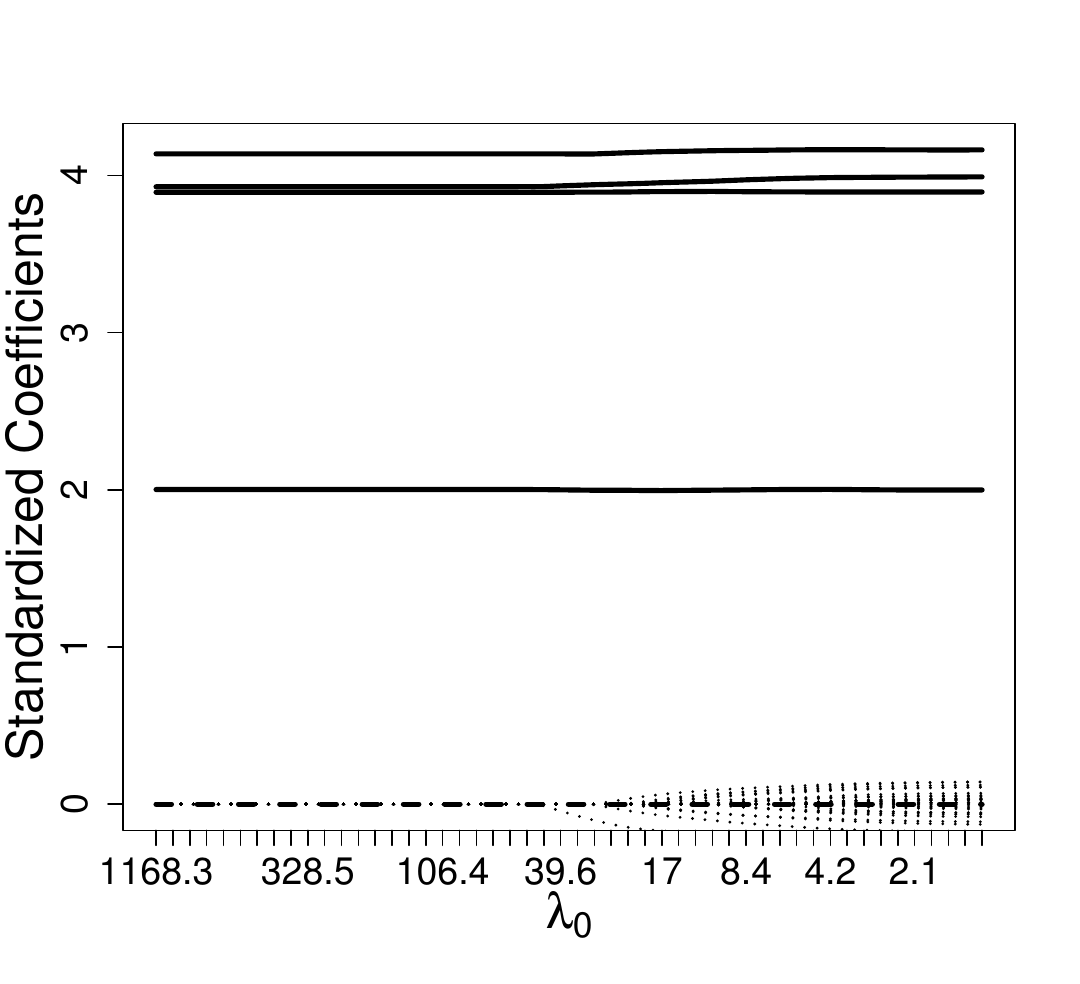}}
\subfigure[SSL, $p=400$]{
\includegraphics[width=.32\textwidth,height=.21\columnwidth]
{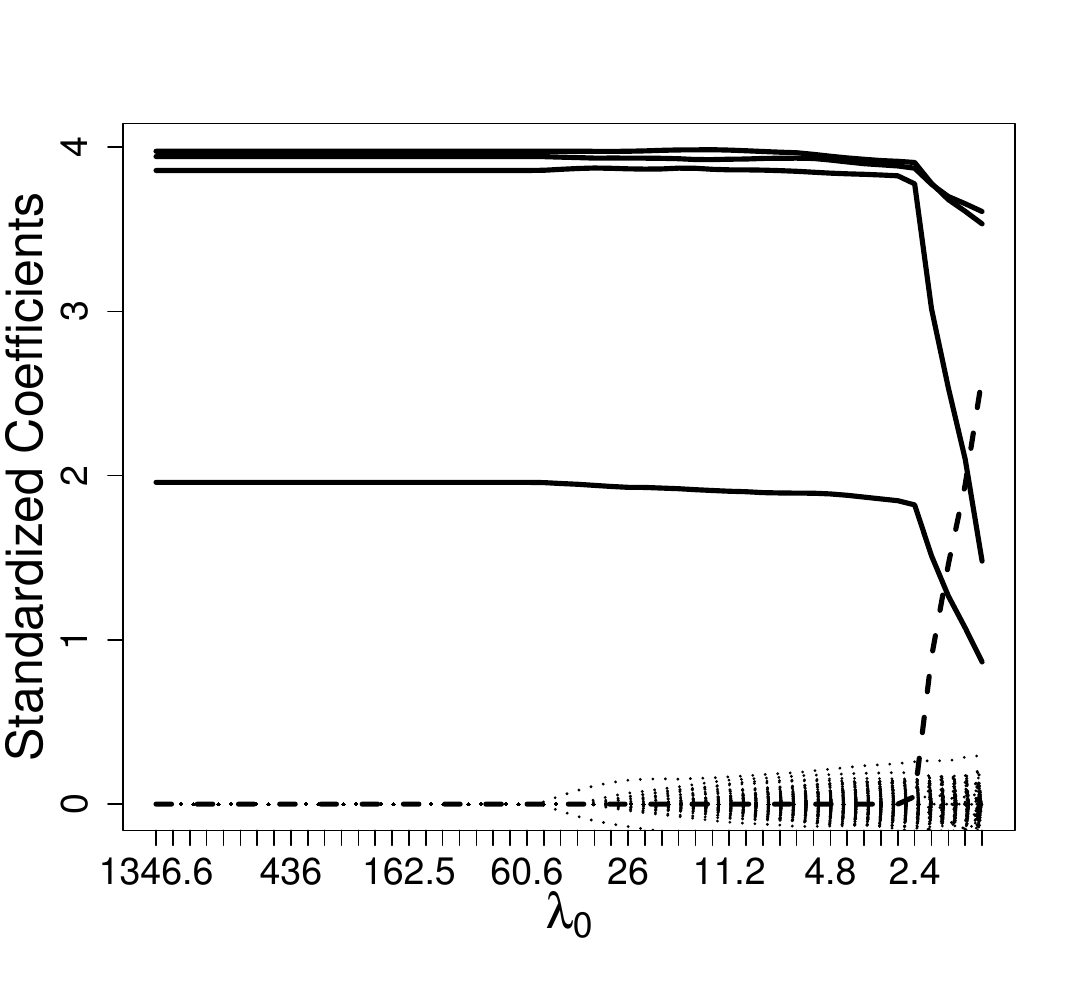}}
\subfigure[SSL, $p=4000$]{
\includegraphics[width=.32\textwidth,height=.21\columnwidth]
{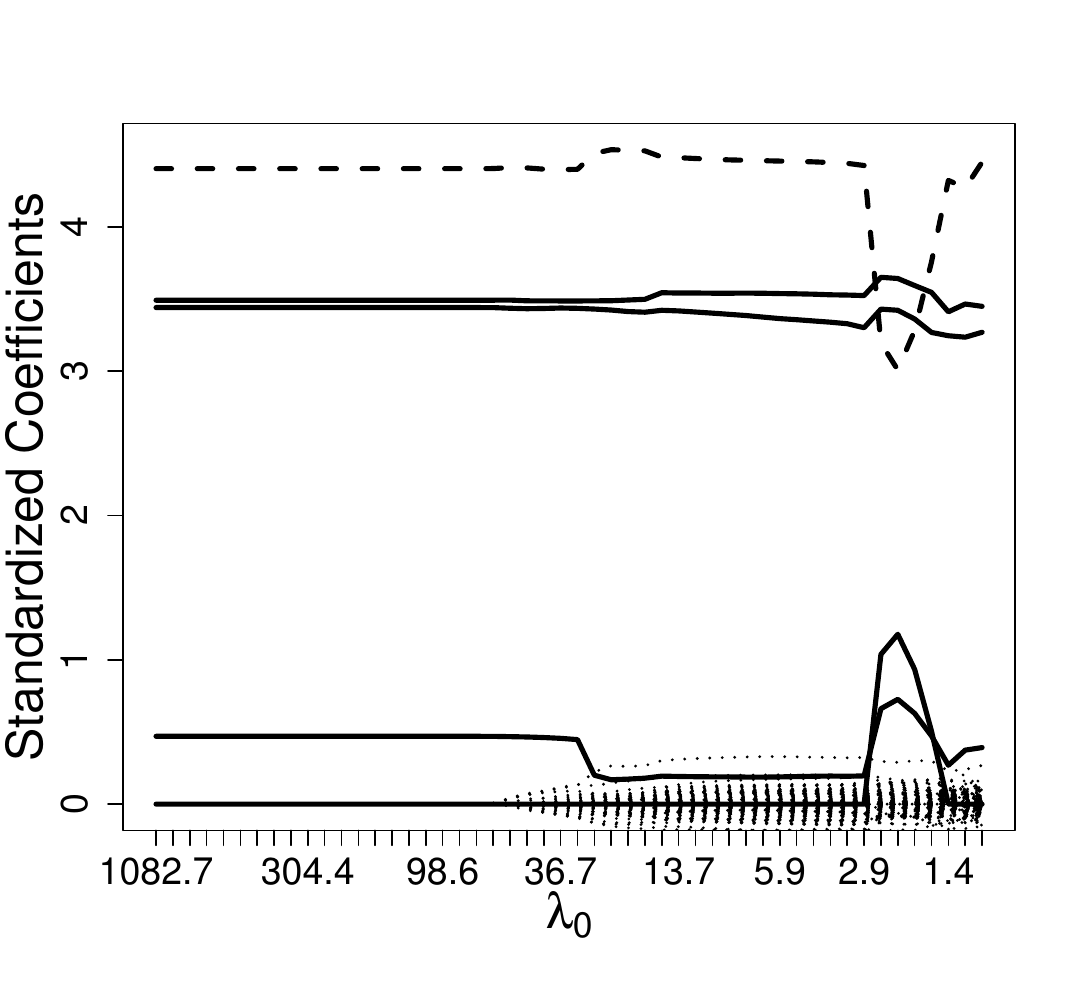}}
\caption{Results for Scenario 1 under three different dimensions. The dashed line corresponds to $X_1$, which is irrelevant; the dotted lines correspond to other irrelevant variables; the solid lines correspond to the relevant variables.}
\label{fig:1}
\end{figure}

\subsection{Performance comparison}\label{section measures}
We consider two settings, $(n, p)=(100, 50)$ and $(n, p)=(100, 200)$.  While cross-validation is not needed for SSL, the tuning parameter for all other methods is selected using 10-fold cross-validation. Each simulation is repeated 100 times. The results are summarized in Tables~\ref{table:1} and \ref{table:2}. It can be seen that MSP uniformly outperforms other methods in both estimation accuracy and model selection.

\begin{figure}[htp]
\centering
\subfigure[$(n,p) = (200,40)$]{
\includegraphics[width=.32\textwidth,height=.35\columnwidth]
{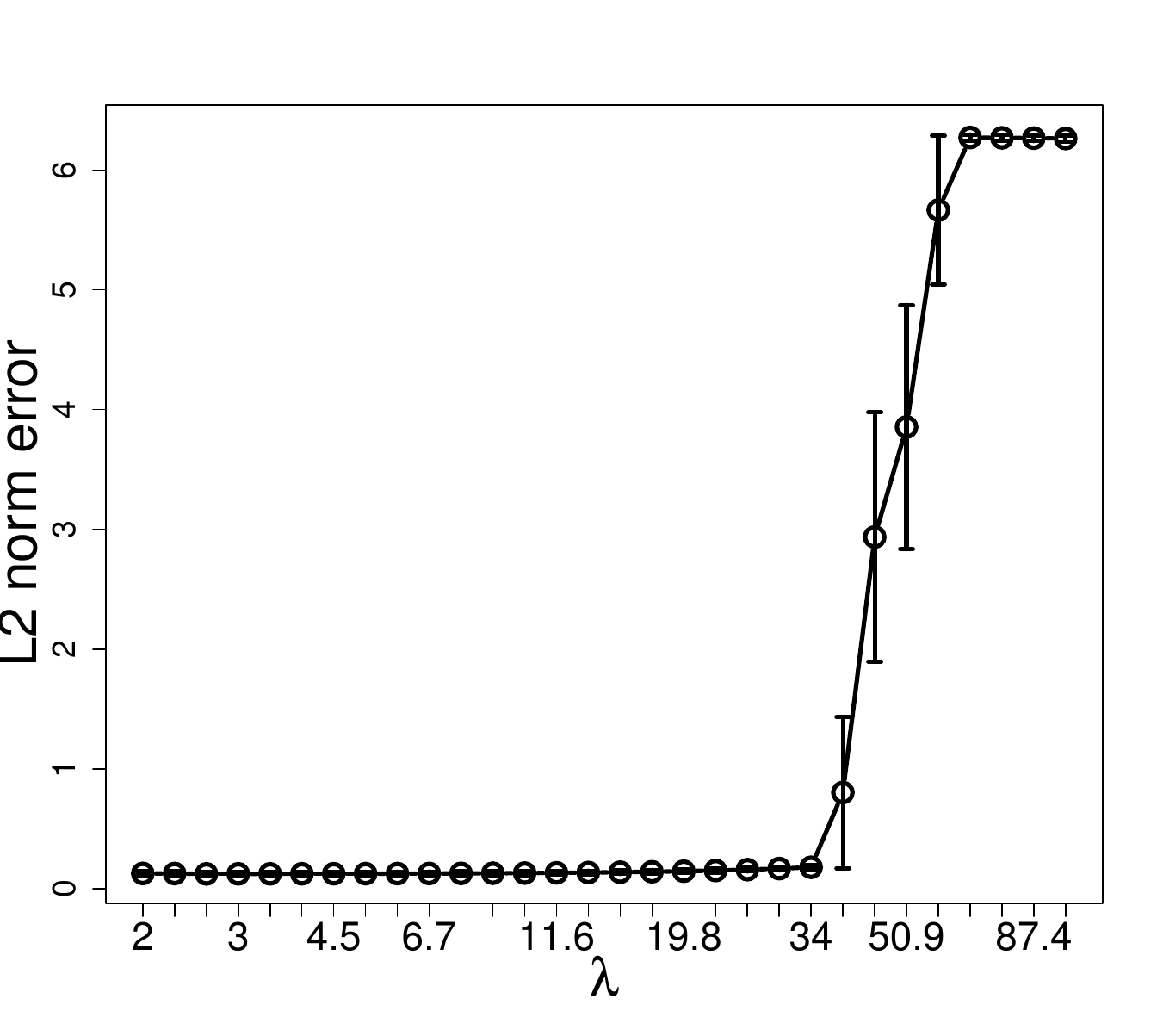}}
\subfigure[$(n,p) = (200,400)$]{
\includegraphics[width=.32\textwidth,height=.35\columnwidth]
{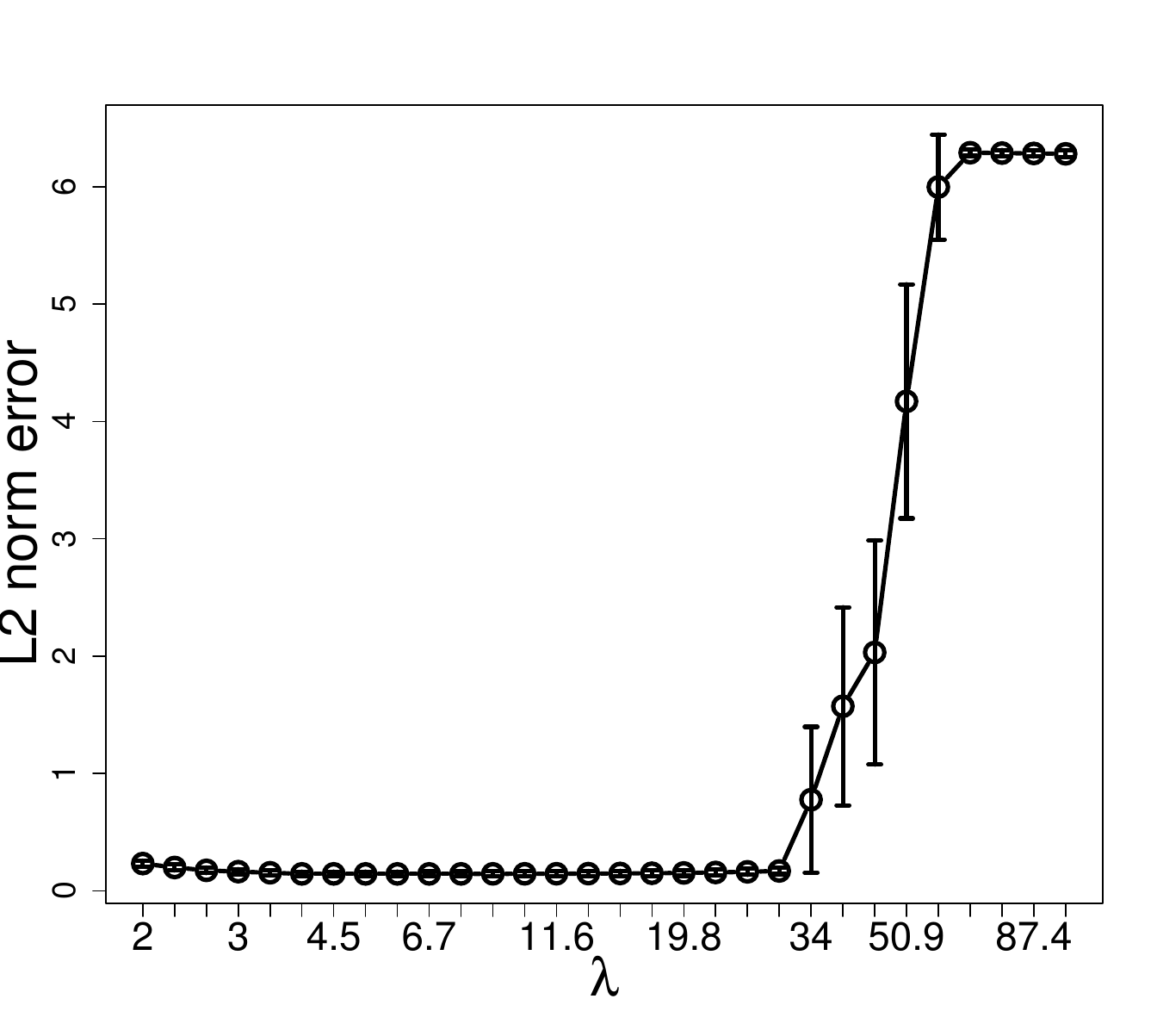}}
\subfigure[$(n,p) = (200,4000)$]{
\includegraphics[width=.32\textwidth,height=.35\columnwidth]
{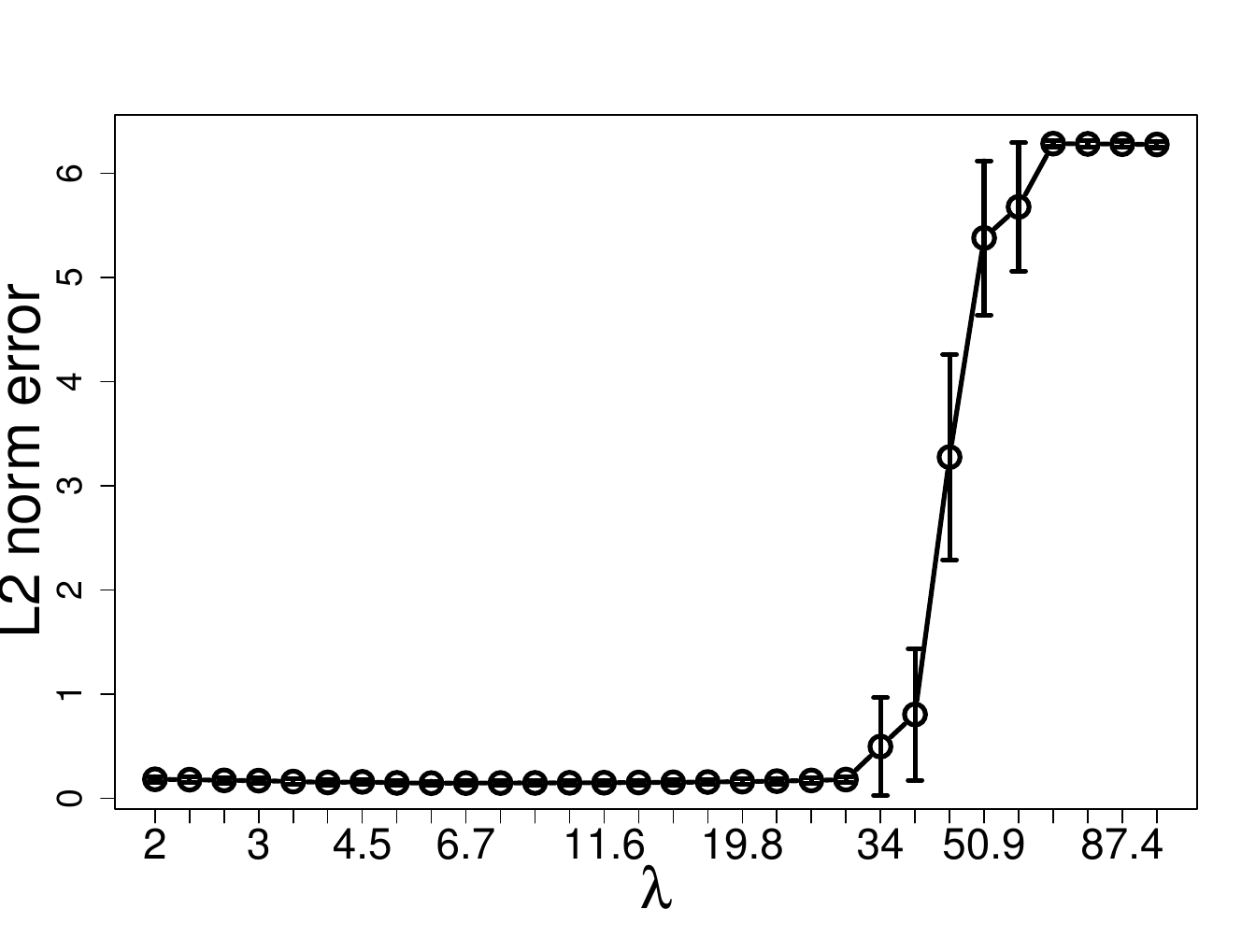}}
\caption{Illustration of the stability of MSP with respect to $\lambda$ under Scenario 1}
\label{fig:lambda}
\end{figure}

Figure~\ref{fig:lambda} shows the estimation error of MSP when $\lambda$ varies under the same simulation setting of Table~\ref{table:2}, i.e. Scenario 1 with $(n,p)=(200,40)$, $(200,400) $ and $(200,4000)$ respectively. As one can see, the estimation error of MSP is low and stable over a range of small $\lambda$ values. The results under Scenario 2 are similar and thus omitted. This implies that cross-validation may not be necessary for MSP in practice; setting $\lambda$ at an appropriately small value often works well, for example, we found $\lambda = (1/5)(n \log n)^{1/2}$ is a reasonable choice after standardizing the design matrix and the response variable.

\begin{table}[htp]
\centering
\small
\def~{\hphantom{0}}
\renewcommand\arraystretch{.6}
\caption{Performance comparison under Scenario 1.}{
\setlength{\tabcolsep}{4 mm}{
\begin{tabular}{lccccc}
\hline\hline
Method 	&	$l_2$-error &		$l_1$-error & NZ	&	FPR	&	TPR	\\
\hline
&& $(n,p)$ &=& (100, 50) &\\
\hline
MSP 	&	0.19 	(	0.07 	)&	0.33 	(	0.12 )&	4.00 	(	0.00 	)&	0.00 	(	0.00 	)&	1.00 	(	0.00 	)\\
Lasso 	&	2.05 	(	0.81 	)&	2.16 	(	0.40 	)&	19.47 	(	2.98 	)&	0.34 	(	0.06 	)&	1.00 	(	0.00 	)\\
MCP	&	6.44 	(	0.12 	)&	3.61 	(	0.06 	)&	7.74 	(	1.22 	)&	0.11 	(	0.02 	)&	0.63 	(	0.13 	)\\
SCAD	&	6.40 	(	0.11 	)&	3.56 	(	0.06 	)&	8.14 	(	1.29 	)&	0.12 	(	0.03 	)&	0.65 	(	0.12 	)\\
SSL 	&	0.31	(	0.61	)&	0.73 	(	1.26 	)&	9.94 	(  3.53 	)&	0.13 	(	0.08 	)&	1.00 	(	0.05 	)\\
ALasso	&	0.38 	(	0.35 	)&	0.79 	(	0.33 	)&	5.56 	(	1.06 	)&	0.03 	(	0.02 	)&	1.00 	(	0.00 	)\\
PLasso	&	1.13 	(	1.29 	)&	1.46 	(	0.63 	)&	8.46 	(	1.45 	)&	0.10 	(	0.03 	)&	0.99 	(	0.05 	)\\
Capped-$l_1$	&	2.05 	(	0.81 	)&	2.16 	(	0.40 	)&	19.47 	(	2.98 	)&	0.34 	(	0.06 	)&	1.00 	(	0.00 	)\\
LLA	&	6.40 	(	0.11 	)&	3.56 	(	0.06 	)&	8.14 	(	1.29 	)&	0.12 	(	0.03 	)&	0.65 	(	0.12 	)\\
\hline
&& $(n,p)$ & =& (100, 200) &\\
\hline
MSP 	&	0.25 	(	0.64 	)&	0.44 	(	1.32 	)&	4.02 	(	0.20 	)&	0.00 	(	0.00 	)&	1.00 	(	0.05 	)\\
Lasso 	&	3.50 	(	0.88 	)&	7.66 	(	1.83 	)&	21.53 	(	3.48 	)&	0.09 	(	0.02 	)&	1.00 	(	0.00 	)\\
MCP	&	6.51 	(	0.70 	)&	14.33 	(	1.65 	)&	18.00 	(	2.66 	)&	0.08 	(	0.01 	)&	0.64 	(	0.13 	)\\
SCAD	&	6.50 	(	0.22 	)&	13.88 	(	0.74 	)&	21.55 	(	3.73 	)&	0.10 	(	0.02 	)&	0.66 	(	0.12 	)\\
SSL 	&	2.77 	(	3.01 	)&	6.41 	(	6.51 	)&	24.88 	(	8.08 	)&	0.11 	(	0.04 	)&	0.88 	(	0.17 	)\\
ALasso	&	1.19 	(	1.36 	)&	2.23 	(	2.58 	)&	5.11 	(	0.96 	)&	0.01 	(	0.01 	)&	1.00 	(	0.03 	)\\
PLasso	&	4.89 	(	2.27 	)&	9.66 	(	4.49 	)&	6.01 	(	1.23 	)&	0.02 	(	0.01 	)&	0.76 	(	0.15 	)\\
Capped-$l_1$	&	3.50 	(	0.88 	)&	7.66 	(	1.83 	)&	21.53 	(	3.48 	)&	0.09 	(	0.02 	)&	1.00 	(	0.00 	)\\
LLA	&	6.50 	(	0.22 	)&	13.88 	(	0.74 	)&	21.55 	(	3.73 	)&	0.10 	(	0.02 	)&	0.66 	(	0.12 	)\\
\hline
\end{tabular}}}
\label{table:1}
\end{table}

\begin{table}[htp]
\def~{\hphantom{0}}
\centering
\small
\renewcommand\arraystretch{.6}
\caption{Performance comparison under Scenario 2. }{
\setlength{\tabcolsep}{4 mm}{
\begin{tabular}{lccccc}
\hline\hline
Method 	&	$l_2$-error &	$l_1$-error & NZ	&	FPR	&	TPR	\\
\hline
&& $(n,p)$ & = & (100, 50) &\\
\hline
MSP 	&	0.22 	(	0.09 	)&	0.37 	(	0.17 	)&	4.00 	(	0.00 	)&	0.00 	(	0.00 	)&	1.00 	(	0.00 	)\\
Lasso 	&	1.82 	(	0.78 	)&	4.91 	(	1.74 	)&	25.87 	(	3.14 	)&	0.48 	(	0.07 	)&	1.00 	(	0.00 	)\\
MCP	&	6.46 	(	0.19 	)&	13.14 	(	0.50 	)&	7.89 	(	1.52 	)&	0.12 	(	0.03 	)&	0.63 	(	0.13 	)\\
SCAD	&	6.43 	(	0.17 	)&	13.44 	(	0.50 	)&	13.60 	(	2.35 	)&	0.24 	(	0.05 	)&	0.69 	(	0.11 	)\\
SSL 	&	0.23 	(	0.09 	)&	0.43 	(	0.19 	)&	5.67 	(	1.84 	)&	0.04 	(	0.04 	)&	1.00 	(	0.00 	)\\
ALasso	&	0.61 	(	0.55 	)&	1.15 	(	1.06 	)&	4.88 	(	0.90 	)&	0.02 	(	0.02 	)&	1.00 	(	0.00 	)\\
PLasso	&	1.14 	(	1.08 	)&	2.63 	(	2.24 	)&	9.18 	(	1.79 	)&	0.11 	(	0.04 	)&	0.99 	(	0.04 	)\\
Capped-$l_1$	&	1.82 	(	0.78 	)&	4.91 	(	1.74 	)&	25.87 	(	3.14 	)&	0.48 	(	0.07 	)&	1.00 	(	0.00 	)\\
LLA	&	6.43 	(	0.17 	)&	13.44 	(	0.50 	)&	13.60 	(	2.35 	)&	0.24 	(	0.05 	)&	0.69 	(	0.11 	)\\
\hline
&& $(n,p)$ &= & (100, 200) &\\
\hline
MSP 	&	0.28 	(	0.63 	)&	0.49 	(	1.28 	)&	4.01 	(	0.10 	)&	0.00 	(	0.00 	)&	1.00 	(	0.05 	)\\
Lasso 	&	2.96 	(	1.11 	)&	2.67 	(	0.43 	)&	33.50 	(	4.38 	)&	0.15 	(	0.02 	)&	1.00 	(	0.03 	)\\
MCP	&	6.48 	(	0.12 	)&	3.62 	(	0.05 	)&	7.67 	(	1.74 	)&	0.03 	(	0.01 	)&	0.55 	(	0.10 	)\\
SCAD	&	6.45 	(	0.12 	)&	3.59 	(	0.05 	)&	8.66 	(	2.24 	)&	0.03 	(	0.01 	)&	0.56 	(	0.10 	)\\
SSL 	&	1.55	(	2.53 	)&	3.24 	(	5.25 	)&	9.73 	(	4.32 	)&	0.03 	(	0.02 	)&	0.93 	(	0.16	)\\
ALasso	&	0.89 	(	1.28 	)&	1.14 	(	0.71 	)&	6.57 	(	1.77 	)&	0.01 	(	0.01 	)&	1.00 	(	0.04 	)\\
PLasso	&	2.01 	(	2.15 	)&	1.95 	(	0.93 	)&	10.43 	(	2.29 	)&	0.03 	(	0.01 	)&	0.95 	(	0.11 	)\\
Capped-$l_1$	&	2.96 	(	1.11 	)&	2.67 	(	0.43 	)&	33.50 	(	4.38 	)&	0.15 	(	0.02 	)&	1.00 	(	0.03 	)\\
LLA	&	6.45 	(	0.12 	)&	3.59 	(	0.05 	)&	8.66 	(	2.24 	)&	0.03 	(	0.01 	)&	0.56 	(	0.10 	)\\
\hline
\end{tabular}}}
\label{table:2}
\end{table}

\subsection{Computational cost}
To compare the computational cost of different methods, we considering the following settings: $n = (100,1000)$ and $p = (100,1000,10000, 100000)$ under Scenario 1. Each running time involves 100 different $\lambda$ values covering a wide range.  For MSP, Lasso, ALasso, LLA, Cappled-$ l_1 $ and MSA, we used the R glmnet package \citep{friedman2010regularization}; for SSL, we used the SSLASSO package \citep{rovckova2018spike}, and for SCAD and MCP, we used the R ncvreg package \citep{breheny2011coordinate}.

Table~\ref{table:3} summarizes the results.  As one can see, the computational cost of MSP is in general larger than that of Lasso, but becomes more comparable as both $n$ and $p$ increase.  In comparison with non-convex one-step methods, including MCP and SCAD, MSP is slower when $n$ and $p$ are small, but faster when $n$ and $p$ are large.  Further, the computational cost of MSP is much lower than that of SSL and those of other multi(two)-step methods, including the Adaptive Lasso, LLA, Capped-$l_1$ and MSA; this is because the MSP only deals with the high-dimensional data in the first step, while other methods deal with the entire data set in every step.

\begin{table}[!htp]
\centering
\small
\def~{\hphantom{0}}
\renewcommand\arraystretch{.6}
\caption{Comparison of average running time in seconds}{
\setlength{\tabcolsep}{1.9 mm}{
\begin{tabular}{lcccccccccc}
\hline\hline
$(n,p)$ &	&	Lasso	&	MCP	&	SCAD	&	MSP	&	ALasso	&	SSL	&	LLA	&	Capped-$l_1$	&	MSA	\\
\hline
($ 10^2 $, $ 10^2 $) &	mean &	0.01 	&	0.05 	&	0.04 	&	0.71 	&	0.28 	&	0.78 	&	1.02 	&	1.46 	&	1.48 	\\
&	sd &	0.01 	&	0.01 	&	0.01 	&	0.02 	&	0.03 	&	0.09 	&	0.41 	&	0.26 	&	0.24 	\\
($ 10^2 $, $ 10^3 $) &	mean &	0.02 	&	0.04 	&	0.07 	&	0.58 	&	0.89 	&	0.77 	&	2.01 	&	4.64 	&	4.04 	\\
&	sd &	0.01 	&	0.01 	&	0.01 	&	0.02 	&	0.07 	&	0.42 	&	0.14 	&	0.75 	&	0.87 	\\
($ 10^2 $, $ 10^4 $) &	mean &	0.25 	&	0.52 	&	0.38 	&	0.59 	&	7.67 	&	4.10 	&	16.47 	&	21.04 	&	18.02 	\\
&	sd &	0.02 	&	0.04 	&	0.03 	&	0.04 	&	0.42 	&	1.58 	&	1.44 	&	2.65 	&	3.76 	\\
($ 10^3 $, $ 10^3 $) &	mean &	0.33 	&	4.08 	&	4.83 	&	2.03 	&	7.31 	&	76.15 	&	95.76 	&	216.24 	&	24.16 	\\
&	sd &	0.02 	&	0.60 	&	0.58 	&	0.09 	&	0.51 	&	3.57 	&	6.13 	&	53.04 	&	4.84 	\\
($ 10^3 $, $ 10^4 $) &	mean &	1.21 	&	7.79 	&	4.28 	&	3.77 	&	43.87 	&	81.75 	&	282.58 	&	388.00 	&	148.18 	\\
&	sd &	0.17 	&	0.70 	&	0.29 	&	0.17 	&	1.67 	&	17.72 	&	36.03 	&	46.56 	&	6.15 	\\
($ 10^3 $, $ 10^5 $) &	mean &	10.96 	&	33.22 	&	32.30 	&	14.48 	&	644.87 	&	302.12 	&	1387 	&	1508 	&	1535 	\\
&	sd &	0.08 	&	3.30 	&	0.22 	&	0.23 	&	29.73 	&	86.47 	&	126 	&	196 	&	224 	\\
\hline
\end{tabular}}}
\label{table:3}
\end{table}

\section{Empirical analysis: index tracking}
In this section, we apply the proposed method to the important and useful index tracking problem in financial modeling.  Roughly speaking, index tracking aims to replicate the movement of a financial index using a small set of financial assets, e.g. stocks, and is the core of the index fund.
This is a high dimensional data modeling problem as the number of stocks that one can choose from is often on the order of hundreds or thousands, while the number of observations (days) is on the order of tens or hundreds. Further, due to transactional cost, one only wishes to select a few rather than many stocks (i.e. a sparse model) to mimic the behavior of the index.

We consider the S\&P500 index and the following model:
$ y_t = \sum_{j=1}^p \beta_j x_{jt} + \epsilon_t, $
where $y_t$ denotes the return of the S\&P500 index on day $t$, $x_{jt}$ denotes the return of stock $j$ on day $t$ and $\beta_j$ is the weight of stock $j$. We consider 19 rolling periods from January 2016 till December 2017 and divide each period into training (=100 days) and testing (=20 days) parts. The training period is used to select stocks and estimate the corresponding $\beta_j$'s and then the testing part is used to evaluate the performance.


We compare 4 methods, including MSP, LLA, Capped-$l_1$ and ALasso, as these four methods had better performances in simulation studies. To measure the performance of different methods, we use the tracking error \citep{meade1989index}, which is a standard measure used in the financial industry to assess the performance of tracking.
It is defined as
\[ \text{TrackingError}_{\text{year}}= \sqrt{252} \times \sqrt{ \sum(\text{err}_t - \text{mean}(\text{err}))^2/(T-1) }, \]
where $\text{err}_t=y_t-\hat y_t$, with $y_t$ and $\hat y_t$ being the daily return of the index and the daily return of the constructed index on day $t$ respectively.

\begin{figure}[!htp]
\includegraphics[width=.48\textwidth,height=.36\columnwidth]
{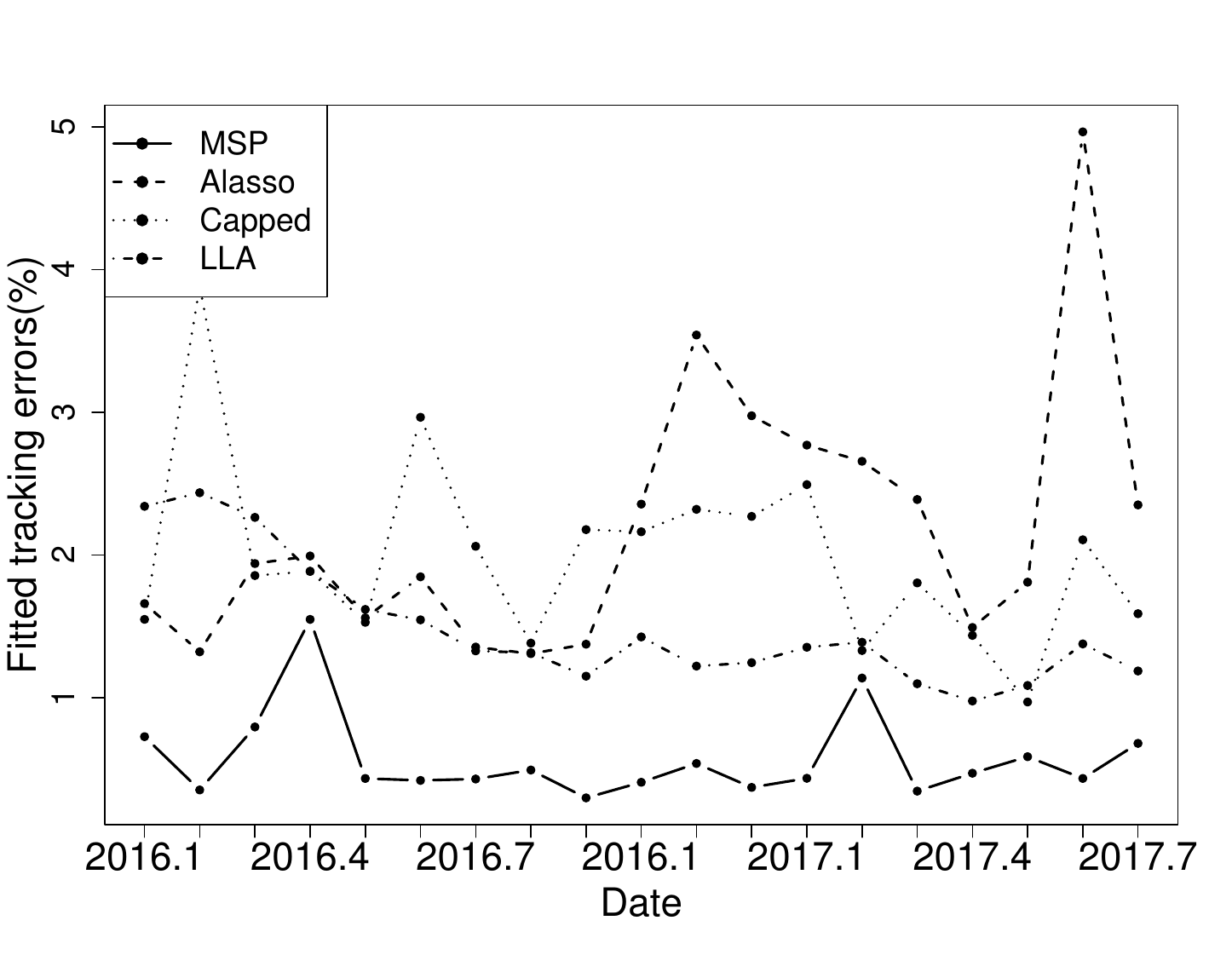}
\includegraphics[width=.48\textwidth,height=.36\columnwidth]
{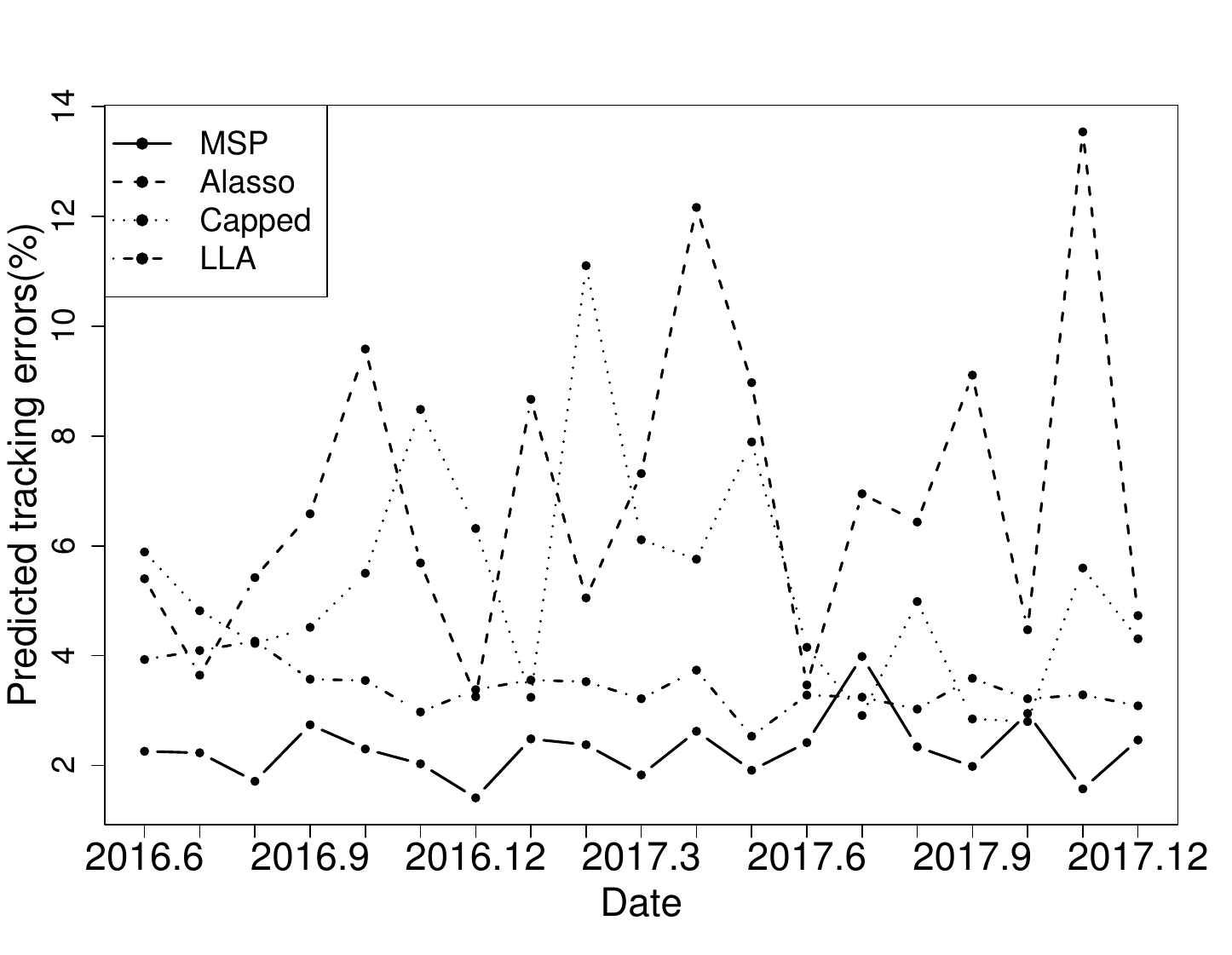}
\caption{Tracking errors for both training (left) and testing (right) sets by different methods}
\label{fig:application}
\end{figure}

We did not use the validation or cross-validation approach to select the tuning parameter; instead, we chose the tuning parameter for each method such that the number of selected stocks is 20, which is often the way done in practice. Figure~\ref{fig:application} shows the 19 tracking errors for both training and testing sets over time. As can be seen, MSP always produces lower and more stable errors than other methods, except for one rolling period.


\bigskip
\begin{center}
{\large\bf SUPPLEMENTARY MATERIAL}
\end{center}

\begin{description}

\item[Title:] Supplementary material for ``MSP: A Multi-step Screening Procedure for Sparse Recovery'. (pdf)

\end{description}


\bibliographystyle{apalike}
\bibliography{reference}
\end{document}